\documentclass[twocolumn,twocolappendix]{aastex701}

\usepackage{array}
\usepackage{booktabs}
\usepackage{tabularx}
\usepackage{dcolumn}
\usepackage{multirow}
\usepackage{amsmath}

\usepackage{array}
\usepackage{booktabs}
\usepackage{tabularx}
\usepackage{graphicx}% Include figure files
\usepackage{dcolumn}% Align table columns on decimal point
\usepackage{bm}% bold math
\usepackage{hyperref}% add hypertext capabilities
\usepackage{xcolor}
\usepackage{multirow}
\usepackage{float}
\usepackage[dvipsnames]{xcolor}

\newcommand{\e}{\epsilon}
\newcommand{\M}{M_{pl}}

 \newcommand{\f}{f_{\mathrm{NL}}}

    \newcommand{\be}{\begin{equation}}
  \newcommand{\ee}{\end{equation}}
    \newcommand{\ba}{\begin{align}}
  \newcommand{\ea}{\end{align}}

\newcommand{\Msun}{M_{\odot}}
\newcommand{\Mpcinv}{ {\rm Mpc}^{-1} }

\newcommand{\MUV}{ M_{\rm UV} }

\newcommand{\sigmaUV}{ \sigma_{\rm UV}}

\begin{document}

\title{A GLIMPSE of Dark Matter: JWST Constraints on Early Structure Formation}

\author[0000-0002-8984-0465]{Julian B. Mu\~noz}
\affiliation{Department of Astronomy, University of Texas at Austin,
2515 Speedway, Austin, TX 78712, USA}
\affiliation{Cosmic Frontier Center, The University of Texas at Austin, Austin, TX 78712, USA}
\affiliation{Weinberg Institute for Theoretical Physics, University of Texas, Austin, TX 78712, USA}
\email[show]{julianbmunoz@utexas.edu}

\author[0000-0002-0302-2577]{John Chisholm}
\affiliation{Department of Astronomy, University of Texas at Austin,
2515 Speedway, Austin, TX 78712, USA}
\affiliation{Cosmic Frontier Center, The University of Texas at Austin, Austin, TX 78712, USA}
\email{chisholm@austin.utexas.edu}

\author[0000-0001-7201-5066]{Seiji Fujimoto}
\affiliation{David A. Dunlap Department of Astronomy and Astrophysics, University of Toronto, 50 St. George Street, Toronto, Ontario, M5S 3H4, Canada}
\affiliation{Dunlap Institute for Astronomy and Astrophysics, 50 St. George Street, Toronto, Ontario, M5S 3H4, Canada}
\email{}

\author[0000-0003-4512-8705]{Tiger Yu-Yang Hsiao}
\affiliation{Department of Astronomy, University of Texas at Austin,
2515 Speedway, Austin, TX 78712, USA}
\affiliation{Cosmic Frontier Center, The University of Texas at Austin, Austin, TX 78712, USA}
\email{}

\author[0000-0001-5538-2614]{Kristen B.~W. McQuinn}
\affiliation{Space Telescope Science Institute, 3700 San Martin Drive, Baltimore, MD 21218, USA}
\affiliation{Department of Physics and Astronomy, Rutgers, The State University of New Jersey, 136 Frelinghuysen Rd, Piscataway, NJ 08854, USA}
\email{}

\author[0000-0002-5554-8896]{Priyamvada Natarajan}
\affiliation{Department of Astronomy, Yale University, New Haven, CT 06511, USA}
\affiliation{Department of Physics, Yale University, New Haven, CT 06511, USA}
\affiliation{Black Hole Initiative, Harvard University, Cambridge, MA 02138, USA}
\email{}

\author[0000-0001-8419-3062]{A. Saldana-Lopez}
\affiliation{Department of Astronomy, Oskar Klein Centre, Stockholm University, 106 91 Stockholm, Sweden}
\email{}

\author[orcid=0000-0002-7570-0824]{Hakim Atek}
\affiliation{Institut d'Astrophysique de Paris, CNRS, Sorbonne Universit\'e, 98bis Boulevard Arago, 75014, Paris, France}
\email{atek@iap.fr}

\author[0000-0001-8104-9751]{Arghyadeep Basu}
\affiliation{Univ Lyon, Univ Lyon1, Ens de Lyon, CNRS, CRAL UMR5574, F-69230, Saint-Genis-Laval, France}
\email{}

\author[0000-0002-4153-053X]{Danielle Berg}
\affiliation{Department of Astronomy, University of Texas at Austin,
2515 Speedway, Austin, TX 78712, USA}
\affiliation{Cosmic Frontier Center, The University of Texas at Austin, Austin, TX 78712, USA}
\email{}

\author[0000-0003-4564-2771]{Ryan Endsley}
\affiliation{Department of Astronomy, University of Texas at Austin,
2515 Speedway, Austin, TX 78712, USA}
\affiliation{Cosmic Frontier Center, The University of Texas at Austin, Austin, TX 78712, USA}
\email{}

\author[orcid=0000-0001-6278-032X]{Lukas J. Furtak}
\affiliation{Department of Astronomy, University of Texas at Austin,
2515 Speedway, Austin, TX 78712, USA}
\affiliation{Cosmic Frontier Center, The University of Texas at Austin, Austin, TX 78712, USA}
\email{furtak@utexas.edu}

\author[0000-0002-3897-6856]{Damien Korber}
\affiliation{Observatoire de Genève, Université de Genève, Chemin Pegasi 51, 1290 Versoix, Switzerland}
\email{damien.korber@protonmail.ch}

\author[0000-0003-3997-5705]{Rohan P. Naidu}
\affiliation{Institute for Astronomy, University of Hawai‘i, 2680 Woodlawn Drive, Honolulu, HI 96822, USA}
\email{}

\author[0000-0002-9651-5716]{Richard Pan}
\affiliation{Department of Physics \& Astronomy, Tufts University, MA 02155, USA}
\email{}

\author[0000-0001-7144-7182]{Daniel Schaerer}
\affiliation{Observatoire de Genève, Université de Genève, Chemin Pegasi 51, 1290 Versoix, Switzerland}
\email{daniel.schaerer@unige.ch}

\begin{abstract}
While we have a solid understanding of how dark matter (DM) behaves during recombination ($z\sim 10^3$) and in the late universe ($z\lesssim$ a few), much less is known about intermediate times, when the first galaxies formed. 
These galaxies provide a powerful handle on DM through structure formation, as halos have to exist to host luminous sources. Here we use recent determinations of the UV luminosity function from the James Webb Space Telescope (JWST) GLIMPSE lensing program, along with legacy Hubble and Subaru data, to test the small-scale behavior of DM during the epoch of reionization ($z\sim 7$).  Marginalizing over astrophysical uncertainties, we find the matter power spectrum to be consistent with the cold-DM prediction down to sub-Mpc scales, with 1$\sigma$ uncertainties of 0.2 dex at $k=3-9 \, \rm Mpc^{-1}$ and 0.5 dex at $k=9-27 \, \rm Mpc^{-1}$, the smallest scales probed at these early times. For thermal-relic warm DM we constrain $m_{\rm WDM} > 3.2\, \rm keV$ (95\% CL), independent of lower-$z$ probes. Our results test small-scale structure formation in a new redshift window and demonstrate the promise of JWST to constrain the nature of DM.  
\end{abstract}

   % UAT terms

\section{Introduction}

While the nature of dark matter (DM) remains a mystery, we know that it is required for galaxies like our own Milky Way to exist. 
The building blocks of structure formation are DM halos, formed from tiny initial
overdensities set in the early Universe that grow under their own
gravity until they collapse and virialize~\citep{Press:1973iz, Bond:1990iw}. 
Galaxies, big and small, form inside these halos~\citep{White:1991mr}, so studying structure formation allows us to test DM through the one interaction we know it has: gravity.

DM is often assumed to be ``cold and collisionless'' and thus able to form structures at all visible scales~\citep[e.g.,][]{Profumo:2006bv}.
Alternatives typically depart from this behavior, for instance suppressing small-scale structures (e.g., in warm, fuzzy, or interacting DM;~\citealt{Bode:2000gq,Hu:2000ke, Cyr-Racine:2015ihg}), or enhancing them (e.g., in models with primordial black holes;~\citealt{Afshordi:2003zb,Inman:2019wvr}). 
Each model leaves its own signature on the matter power spectrum $P(k)$, which quantifies the variance of density fluctuations at each wavenumber $k$ (corresponding to scales $\sim 1/k$ in comoving Mpc). Measuring this quantity across scales and times provides a powerful handle on the nature of DM. 

\begin{figure}
    \centering
    \includegraphics[width=0.9992\linewidth]{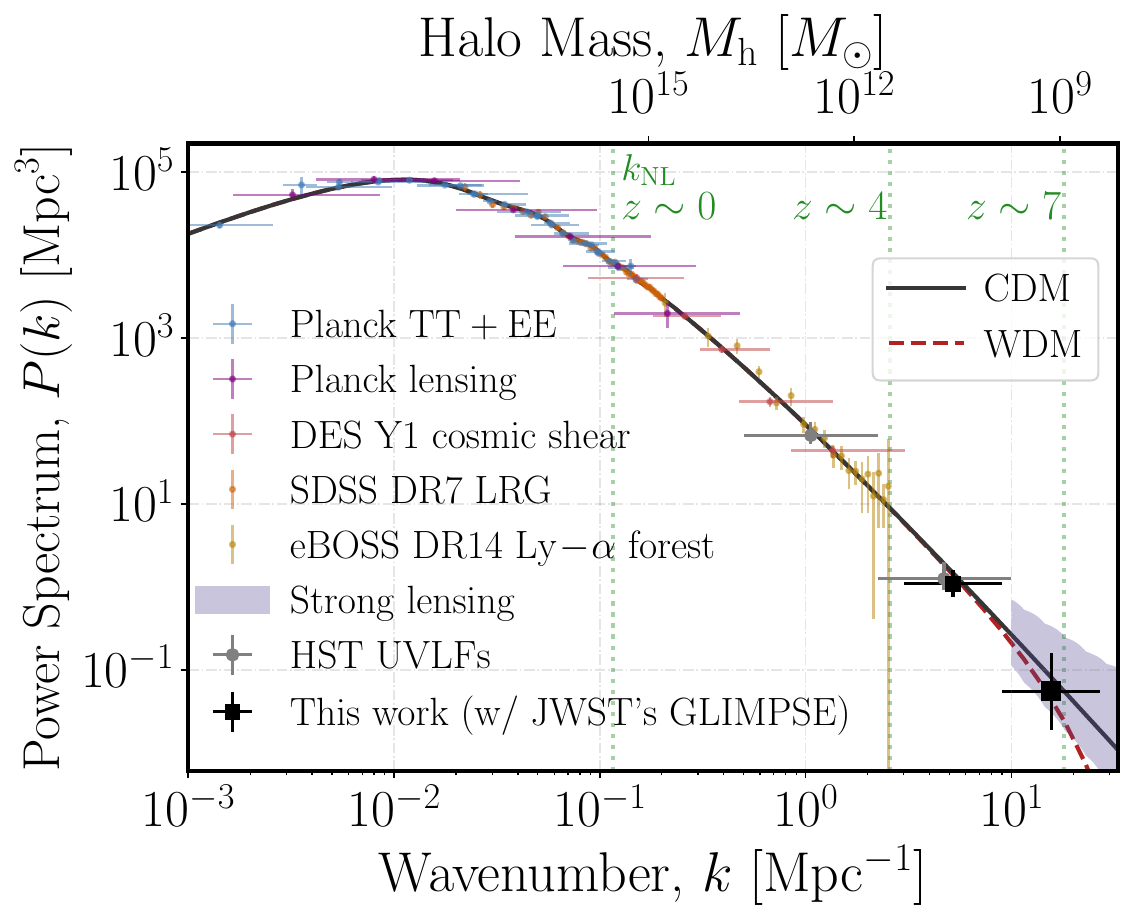}
    \caption{The matter power spectrum $P(k)$ linearly extrapolated to $z=0$. Black line is the CDM prediction and  red dashed is WDM with $m_{\rm WDM}=3.2$ keV (at our 95\% CL).
    Colored circles show measurements from different probes~\citep{Chabanier:2019eai,Planck:2018nkj,DES:2017qwj,Reid:2009xm,SDSS:2017bih}, and vertical green-dotted lines correspond to the non-linear scale (defined as $k_{\rm NL} = 1/R_{\rm NL}$ for the radius $R_{\rm NL}$ that has unit variance) at $z\sim 0,~4$ and $7$. 
    Our results are shown as black squares, which reach comparable scales to strong lensing~\citep[][purple region]{Gilman:2021gkj} but at higher $z$, and smaller scales (higher $k$, smaller halo masses on the top axis) than past UVLF work with HST~\citep[][gray]{Sabti:2021unj}. 
    }
    \label{fig:Pk_result}
\end{figure}

Over the last few decades we have made great progress on measuring $P(k)$, as summarized in Fig.~\ref{fig:Pk_result}. 
At large scales  ($k\lesssim 1\,\Mpcinv$ or separations larger than a few Mpc) the cosmic microwave background (CMB) and large-scale structure (LSS) have reached percent-level precision at early and late times, respectively~\citep{DES:2021wwk}. 
At small scales (sub-Mpc, or $k\gtrsim {\rm a\, few} \,\Mpcinv$), however, we only have late-time constraints, arising from  Milky Way satellites~\citep{DES:2020fxi}, stellar streams~\citep{Banik:2019smi}, lensing substructure in galaxies~\citep{Vegetti:2009cz} and clusters~\citep{Natarajan+2026,Chiang+2026}, 
and the Lyman-$\alpha$
forest~\citep[][which reaches $z\sim 2-5$]{Irsic:2017ixq}.
While powerful, these probes are complicated by billions of years of nonlinear structure formation and accumulated feedback~\citep{Garzilli:2019qki,Bullock:2017xww}. 
As a result, they can set strong constraints on specific DM models or parametrized forms of $P(k)$~\citep[e.g.,][]{Dekker:2024nkb}, rather than measure it scale by scale.

At early times (redshifts $z\sim 5-20$, when the first galaxies formed and reionized the Universe) the matter power spectrum remains largely uncharted. This high-$z$ gap is not just a curiosity. 
Structures form hierarchically in our Universe, from small to big.
For instance, the typical halo hosting $z\sim 7$ faint galaxies has $M_h\sim 10^{9}\,\Msun$, whereas the Milky Way today resides in a $M_h\sim 10^{12}\, \Msun$ halo.
As such, the same small-scale (high-$k$) modes that are deep in the nonlinear regime at $z\sim 0$ were just barely forming structures during reionization, so high-$z$ structures allow us to test these fluctuations in a regime closer to linearity (and with $\lesssim 1$ Gyr of built-up feedback). 
A powerful avenue to access this early Universe is the UV luminosity function (UVLF), which measures the abundance of star-forming galaxies against their luminosity.
Determinations of this quantity using the {\it Hubble} Space Telescope (HST) over the last decade have become increasingly refined~\citep[e.g.,][]{Finkelstein2016,Oesch_2018,Bouwens:2014fua}, yielding constraints on DM models~\citep{Menci:2017nsr,Rudakovskyi:2021jyf,Corasaniti:2016epp,Dayal:2014nva,Schultz:2014eia,Liu:2024edl,Lazare:2025gha,Barron:2025dys}, inflation~\citep{Sabti:2020ser,Parashari:2023cui}, and generic departures from cold DM (CDM) up to $k\sim 10\,\Mpcinv$~\citep[][shown in Fig.~\ref{fig:Pk_result}]{Sabti:2021unj}.
Reaching smaller scales, where DM departures are more readily apparent, requires deeper observations to catch the faint galaxies that live in lower-mass halos.

Here we use UVLFs from the JWST lensing program GLIMPSE~\citep{AtekChisholm26_Glimpse_survey} to measure the small-scale structure of DM.
GLIMPSE exploits the gravitational magnification by the cluster Abell S1063 to detect galaxies  down to an unprecedented depth of $M_{\rm UV}\sim -12$~\citep{Atek_26_UVLFGLimpse}.
%We extend our Zeus21/GALLUMI public pipeline~\citep{Munoz:2023kkg,Cruz:2024fsv} to model UVLFs and biases for non-CDM models at high $z$.
We fit these data, along with legacy HST and Subaru observations at $z\sim 4-8$ to measure the matter power spectrum $P(k)$ at small scales, 
marginalizing over astrophysical uncertainties.
Fig.~\ref{fig:Pk_result} summarizes our results. We constrain the amplitude of $P(k)$ to within $0.2-0.5$ dex of the CDM prediction up to $k\sim 27\,\Mpcinv$, and set a bound of $m_{\rm WDM}>3.2$~keV for warm DM (WDM, at 95\% CL). 
These constraints rely on different modeling and systematics than the CMB, LSS, and small-scale probes, and test a complementary range of times and scales, filling in our understanding of structure formation during the first billion years. 
Throughout this work we assume a {\it Planck} 2018 cosmology~\citep{Planck:2018vyg}, massless neutrinos, and AB magnitudes~\citep{OkeGunnAB} unless otherwise specified.

\section{Fitting the UVLFs}

The UVLF $\Phi_{\rm UV}$ measures the abundance of star-forming galaxies by counting their comoving number density per unit UV magnitude $\MUV$,
\be
\Phi_{\rm UV}(M_{\rm UV}) = \int d M_{\rm h} \dfrac{dn}{dM_{\rm h}} (M_h) P(M_{\rm UV} |M_{\rm h}).
\ee
Modeling the UVLF requires two ingredients: a halo-galaxy connection $P(M_{\rm UV} |M_{\rm h})$, which encapsulates the astrophysics,
and the halo mass function (HMF) $dn/dM_{\rm h}$, which depends on cosmology.
Let us describe each in turn.

\subsection{The halo-galaxy connection}

How galaxies populate DM halos is still an open question in high-$z$ astrophysics~\citep{Wechsler:2018pic}, as it encodes information on feedback, accretion, and stellar physics.  
We follow an analytic approach where we place galaxies with $\MUV$ in halos of mass $M_{\rm h}$ via a PDF~\citep{Sabti:2021xvh}
\be
P(M_{\rm UV} |M_{\rm h}) = \frac{e^{-(\MUV - \overline M_{\rm UV})^2/(2\sigmaUV^2)}}{\sqrt{2\pi\sigmaUV^2}},
\label{eq:PDF_MUV_Mh}
\ee
where both the centroid $\overline M_{\rm UV}$ and the rms $\sigma_{\rm UV}$ depend on $M_h$ and $z$ through free parameters  described below.
This simple halo-galaxy connection ignores satellites (though it can be enhanced to include them; e.g., \citealt{Berlind:2001xk}), which is expected to be accurate for the scales and redshifts of interest~\citep{Harikane2021,Bhowmick:2018waq} and can be generalized beyond Gaussianity~\citep{Hegde:2025mgj}.

The UVLF tracks star formation, as
UV light is emitted by young ($\lesssim 100$ Myr) stars.
Therefore, we can relate the UV luminosity (and thus $\MUV$) of a galaxy to its recent star-formation rate $\dot M_\star$ through 
\be
L_{\rm UV} = \dot M_\star/\kappa_{\rm UV}   = f_\star f_b \dot M_h/\kappa_{\rm UV},
\ee
where $\kappa_{\rm UV}$ is a SFR-UV conversion factor (set to $1.15 \times 10^{-28} \,\rm \Msun yr^{-1}/( erg \,s^{-1} \, Hz^{-1})$ for standard PopII stars with a Salpeter IMF and constant SFR, \citealt{Madau:2014bja}), $\dot M_h$ is the mass accretion rate (from the exponential fit of \citealt{Dekel:2013uaa}), $f_b\approx 0.16$ is the baryon fraction, and $f_\star$ is the star-formation efficiency (SFE) which depends on $M_h$ and $z$. 
Given that we only measure the product of these factors, any uncertainty on the overall accretion rate or SFR-UV conversion (e.g., through the initial mass function, star-formation histories, or dust geometry) can be encoded in the average SFE, so we will focus here on variations of this quantity.

We model the SFE $f_\star(M_h,z)$ flexibly to capture the effect of feedback and marginalize over it. 
Our current understanding of early-Universe galaxy formation suggests that the star formation is less efficient at the heavy end due to AGN feedback~\citep{Silk:1997xw}, and at the low-mass end from a combination of supernova feedback, radiative heating, and reionization~\citep{Efstathiou92_feedback,Hopkins:2013vha,Borrow:2022djx}.  
Based on empirical and semi-analytic models~\citep{Mason:2015cna,Behroozi:2019kql}, we take a double-power law functional form for the SFE:
\be
f_\star (M_h) = \dfrac{2 \, \epsilon_\star}{(M_h/M_p)^{-\alpha_\star} + (M_h/M_p)^{-\beta_\star}},
\ee
with four free parameters: an SFE amplitude $\epsilon_\star$, pivot mass $M_p$, and suppression indices for the faint and bright ends $\alpha_\star \geq 0$ and $\beta_\star \leq 0$, respectively. 
These can all potentially vary with redshift, and we will calibrate them against observations of UVLFs and clustering.

We include dust following a parametrized prescription. 
Dust will both dim and redden galaxies, creating a correlation between attenuation $A_{\rm UV}$
and UV-slope $\beta_{\rm UV}$~\citep{Meurer:1999jj}:
\be
A_{\rm UV} = C_{1,\rm dust} (\beta_{\rm UV} - \beta_{0,\rm dust}),
\ee
where we take $C_{1,\rm dust}=1.5$ and $\beta_{0,\rm dust}=-2.23$ from \cite{Bouwens16xiion} for our fiducial analysis, but will also show results when varying these parameters (which do not alter our constraints, as dust mainly affects the massive end). 
We apply dust (keeping $A_{\rm UV} \geq 0$) with this relation to our model before comparing with data, given the mean $\beta_{\rm UV}-\MUV$ relation from \citet[][see also~\citealt{Jecmen26_betaUV}]{Bouwens:2013hxa} and assuming a scatter of $\sigma(\beta_{\rm UV})=0.34$~\cite[][which, multiplied by $C_{1,\rm dust}$, adds in quadrature to the input $\sigma_{\rm UV}$]{Tacchella:2012ih}.

Finally, one of the key discoveries of JWST is the prevalence of time variability or ``burstiness'' in star formation at high $z$~\citep{Endsley25_burstiness, Sun25_burstiness}, which broadens the halo-galaxy connection as a higher $\sigma_{\rm UV}$~\citep{Mirocha_UVLFs2023,Shen:2023cva, Gelli24_burstinessM}. 
Recent work shows that burstiness increases toward lower halo masses~\citep{Munoz:2026gxp} and at earlier times~\citep{Ciesla_24_burstiness}.
To remain robust to this ingredient we allow the UV scatter $\sigma_{\rm UV}$ to freely vary against halo mass and redshift.

\subsection{The HMF and cosmology}

The second ingredient that determines the UVLF is the abundance of DM halos that can host galaxies, which we quantify through the HMF $dn/dM_h$.  
The HMF is obtained as a function of the (matter) density variance
\be
\sigma^2(M_h) = \int \frac{d^3k}{(2\pi)^3} P(k) |W(k;M_h)|^2, 
\ee
where $W(k;M_h)$ is a window function that encodes how fluctuations of wavenumber $k$ contribute to form halos of mass $M_h$. 
This equation shows that the HMF is a powerful (albeit nonlinear) tracer of DM clustering through $P(k)$. 
As an example, models such as warm or fuzzy DM suppress $P(k)$ at high $k$, producing a plateau in the matter variance $\sigma^2(M_h)$ at small $M_h$, and therefore a sharp reduction in the HMF compared to CDM~\citep[e.g.,][]{Schneider:2011yu,Marsh:2013ywa}.

The HMF has been extensively simulated at low-$z$ both in CDM and non-CDM universes~\citep[e.g.,][]{Tinker:2008ff,Angulo:2013sza,Vogelsberger:2015gpr,Benson:2026kdp}. 
From these studies we have learned that selecting the right window function $W$ is key~\citep{Leo:2018odn}.  
In the usual CDM scenario the power spectrum $P(k)$ is smooth and the window function can simply be a top-hat with a radius that encloses the mass $M_h$ of the halo.
For beyond-CDM models, however, the power-spectrum can have features (suppression, enhancement, or even oscillations), so the traditional top-hat window function fails. Instead, \citet{Benson:2012su} advocate for a
``sharp-k'' window function (i.e., a Heaviside theta function in $k$ space), cutting at $k = c (4\pi \rho_m/3M_h)^{1/3}$, where $\rho_m$ is the matter density and $c=2.5$ is a calibration constant, which takes the place of the Sheth-Tormen  peak-height rescaling parameter ($a_{\rm ST}\approx 0.7$ for a top-hat window, usually set to unity for sharp-$k$).

While there is consensus on the HMF  at $z\sim 0$, much less is known at high $z$. 
Only recently have CDM simulations begun to  converge on the HMF at $z\sim 5-10$, thanks to suites combining large volumes and high resolution with consistent initial conditions and halo finders~\citep{Yung:2023sib}, but there are comparatively few high-$z$ simulations beyond CDM~\citep[e.g.,][]{Schive:2015kza, Schneider:2018xba, Bohr:2021bdm}.  
We control for this potential systematic in three ways. 
First, high-$z$ simulations find $a_{\rm ST}$ values in the $0.9-1$ range for sharp-$k$ window functions~\citep{Barron:2025dys}, so we will vary this parameter in our MCMCs in that range. Second, some works argue for $c=2.7$ rather than 2.5~\citep{Schneider:2013ria}, so we will show results for this value as well.   
Finally, we will re-calibrate the HMFs and biases to the CDM case. 
That is, a non-CDM scenario will have an HMF
\be
\frac{dn_{\rm nCDM}}{d M_h} = \frac{dn^{\rm Yung+}_{\rm CDM}}{d M_h}\times \frac{dn^{\rm ST}_{\rm nCDM}/dM_h }{ dn^{\rm ST}_{\rm CDM}/dM_h }
\ee
where $dn^{\rm ST}/dM_h$ (along with $\sigma(M_h)$, both for CDM and non-CDM) is computed with a ``standard'' \citet{Sheth:1999su} HMF given a sharp-$k$ window function and input $a_{\rm ST}$, and $dn^{\rm Yung+}/dM_h$ is the CDM fit from~\citet{Yung:2023sib}.
This procedure ensures that a CDM $P(k)$ input recovers the correct HMF \cite[and bias, from][based on \citealt{Tinker:2010my}]{Wang:2026hwx} regardless of window function. 

\begin{figure}
    \centering
    \includegraphics[width=0.98\linewidth]{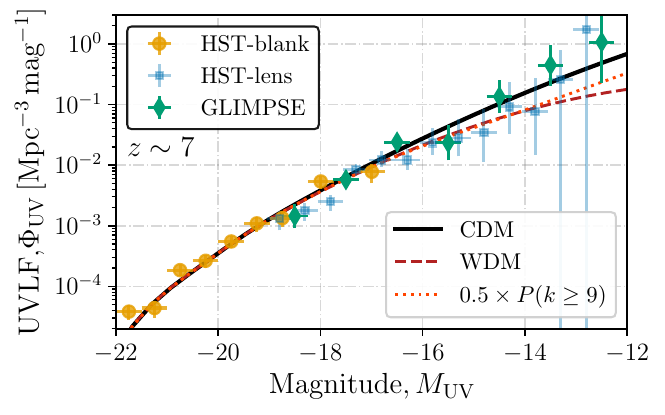}
    \caption{UVLF at $z\sim 7$ predicted within CDM (solid), 
    WDM (with $m_{\rm WDM} = 3.2$ keV, red dashed), and a model with a 2$\times$ reduction in power for $k\geq 9\,\Mpcinv$ (orange dotted), alongside data from HST-blank fields (yellow circles), HST-lens (blue squares), and JWST's GLIMPSE (green diamonds). 
    The depth of GLIMPSE allows us to better constrain small-scale structure and thus deviations from CDM. 
    }
    \label{fig:UVLF_z7}
\end{figure}

The model above allows us to connect any input DM model or $P(k)$ to a HMF and then, by ``dressing it'' with a halo-galaxy connection, to predict a UVLF. 
As an example, Fig.~\ref{fig:UVLF_z7} shows the UVLF at $z\sim 7$ for CDM and a WDM case (with $m_{\rm WDM} = 3.2$ keV).
The CDM UVLF grows towards the faint end along with the abundance of small-mass halos able to host faint galaxies. The WDM scenario, on the other hand, shows a decline in the galaxy abundance at the faint end (similar to halving the power $P(k)$ for $k\geq 9\,\Mpcinv$, shown for reference). 
The faint end of the UVLF can, therefore, test the small-scale behavior of DM.

\subsection{The UVLF and likelihood}

Broadly speaking, UVLF determinations come in two flavors: blank and lensed. 
The former are obtained through ``random'' telescope pointings, whereas the latter are chosen near low-$z$ massive clusters that magnify high-$z$ sources behind them. 
We will combine these data, choosing compilations that overlap minimally to avoid double counting: 
the {\bf HST-blank} UVLFs compiled in~\citet{Bouwens21}, the {\bf HST-lens} from~\citet[][see also~\citealt{FinkelsteinBagley22_UVLF,Atek:2018nsc} for alternate determinations]{Bouwens:2022ojz}, 
the Subaru/{\bf HSC} LFs and biases from~\citet{Harikane2021}, and the new lensed data from JWST's {\bf GLIMPSE}~\citep[][overlapping HST-lens in only one of its six clusters, Abell S1063]{Atek_26_UVLFGLimpse}. 
We restrict ourselves to $z\sim 4-8$ and $\MUV > -22$ as dust and AGN become more prevalent at lower $z$ and brighter magnitudes~\citep{Bowler:2014vra,Ono:2017wjz}, and the UVLF at $z\gtrsim 9$ may need new modeling ingredients~\citep{Adamo25_reviewJWST3yrs}, so we remain within the better-understood $z\leq 8$ regime to obtain robust limits.

Fig.~\ref{fig:UVLF_z7} shows the $z\sim 7$ UVLF data, where {HST-blank} has the best measurements at the bright end, {HST-lens} at intermediate $\MUV\sim -17$, and the new {GLIMPSE} data at the faint end ($\MUV\gtrsim -16$). 
Together, they show the UVLF rises steeply towards the faint end in agreement with the prediction of CDM with our fiducial halo-galaxy connection. 
Given these data, we build a likelihood as in \citet{Sabti:2021xvh} by assuming Gaussian uncorrelated errors between bins (appropriate for noise-limited samples), with a minimum 20\% error to account for cosmic variance \citep[][which can affect nearby bins coherently, but we leave off-diagonal covariances for future work]{Trapp_2020}.

We also fit bias data from Subaru/HSC at $z\sim 4-6$ with a Gaussian likelihood, which is summed to that of the UVLF (as summarized in Appendix E of \citealt{Munoz:2026gxp}, also with a minimum 20\% uncertainty, in this case to account for possible satellite/central mis-modeling).   
The galaxy bias measures the typical mass of the DM host halos, so it will help break degeneracies between parameters, such as SFE and burstiness~\citep{Munoz:2023cup}.

% Then, the log-likelihood is
% \be
% -\log \mathcal L = \sum \dfrac{(\Phi_{\rm UV}^{\rm meas.} - \Phi_{\rm UV}^{\rm model})^2}{2 \sigma_{\Phi}^2}
% \ee
% summing over $\MUV$ and $z$, where the UVLF error $\sigma_{\Phi}$ has 

\section{Results}

Before setting DM constraints, we perform an MCMC search over all the astrophysical parameters to identify a ``fiducial'' model able to explain our data within standard CDM.
We find that a nearly $z$-independent SFE fits the data well (as shown by~\citealt{Tacchella:2018qny}), and allowing either $\epsilon_\star$ or $M_p$ to vary with redshift provides a slightly better fit ($\Delta \chi^2 \approx 4$ for 1 parameter), so we keep the former as redshift dependent (and find consistent results with the alternative choice).
As for burstiness, we allow  $\sigmaUV$ to vary linearly against (log)$M_h$ and $z$. 
This will be our fiducial model, with 5 free SFE parameters and 3 for burstiness, whose ranges and priors are summarized in Appendix~\ref{app:parameters}.
It can fit the $z\sim 4-8$ UVLFs with $\chi^2/(\rm d.o.f.) = 0.86$, indicating slightly overestimated error-bars.
We will use this model to perform two different searches for DM.

\subsection{Constraining the Matter Power Spectrum}

Each DM model will give rise to a different matter power spectrum $P(k)$. 
Rather than searching for a specific deviation template, we first perform a parametric (DM-model agnostic) search for departures from CDM by allowing the power to vary separately in $k$ bins as in~\citet{Sabti:2021unj}. 
That is, we set
\be
P(k) = 
\begin{cases}
P_{\text{CDM}}(k) & \text{if } k < 3\,\text{Mpc}^{-1} \\
a_1 P_{\text{CDM}}(k) & \text{if } 3 \leq k < 9\,\text{Mpc}^{-1} \\
a_2 P_{\text{CDM}}(k) & \text{if } 9 \leq k < 27\,\text{Mpc}^{-1} \\
a_3 P_{\text{CDM}}(k) & \text{if } k \geq 27\,\text{Mpc}^{-1} 
\end{cases}\, ,
\ee
where we have chosen the $k$ ranges to follow the sensitivity of our UVLF data (as detailed in Appendix~\ref{app:krange}), and the 
number of bins to resolve different scales while minimizing covariances.

Note that we assume the $k < 3\,\Mpcinv$ power spectrum follows CDM, as it is tightly constrained by other observables (and varying $A_s$ or $\sigma_8$ within CMB uncertainties changes the power spectrum by $\lesssim 0.01$ dex, far below the precision we will reach). 
Moreover, the last bin is chosen to be beyond measurable $k$, so we can set its amplitude to unity safely ($a_3=1$).
Thus, in this first analysis we vary the amplitude of the matter power spectrum from $k=3-9\,\Mpcinv$ and $9-27\,\Mpcinv$ by adding three free parameters to our likelihood: $\log_{10} a_1$, $\log_{10} a_2$, and $a_{\rm ST}$, all with flat priors (the latter over $[0.9,1]$, the other two over $[-1.5,1.5]$).

We find agreement with CDM at all scales we probe. 
The $k=3-9\,\Mpcinv$ bin has an amplitude $\log_{10} a_1 = -0.16^{+0.17}_{-0.16}$, 
corresponding to a $\approx$ 40\% measurement of the power spectrum.  
At smaller scales ($k=9-27\,\Mpcinv$) we find $\log_{10} a_2= -0.19^{+0.45}_{-0.46}$, with a half-dex or $\mathcal O(1)$ uncertainty. 
These results reach both better precision ($\approx 2\times$ at $k\lesssim 10\,\Mpcinv$) and smaller scales ($\approx 3\times$ larger $k$)  than earlier UVLF work in \citet{Sabti:2021unj}, partly due to the deeper GLIMPSE data (which provide fainter and likely lower-mass objects) and partly due to fitting Subaru/HSC biases (which break degeneracies between parameters). 
We find that the posteriors of the two amplitudes $a_1$ and $a_2$ are correlated ($r\approx 0.7$), and fixing the former (e.g., from other cosmic probes at $k\sim 3-9\,\Mpcinv$) would reduce the uncertainty on the latter from 0.5 to 0.3 dex.

Fig.~\ref{fig:Pk_result} places our measurements in context alongside other $P(k)$ constraints, showing no significant departures from CDM up to $k\approx 27\,\Mpcinv$.  
The next subsection will discuss the implications of this result in the context of WDM as an example of $P(k)$ suppression, but we note that this constraint also applies to models where small-scale power is enhanced~\citep[e.g.,][]{deKruijf:2024voc}.
These $P(k)$ measurements are far from the percent level, yet they reach earlier cosmic times ($z\sim 4-8$) than other probes of small-scale structure, extending our understanding of DM and structure formation to the epoch of reionization. 

\subsection{WDM Constraints}

The model-agnostic search above is not optimal for specific DM models where $P(k)$ is known.
To illustrate this, we search for a thermal WDM relic, a common benchmark alternative to CDM where power is suppressed at small scales due to the free-streaming nature of the DM particles.
This reduces the abundance of small-mass halos and therefore of faint galaxies, as shown in Fig.~\ref{fig:UVLF_z7}.
However, this signature reduction can be mimicked by astrophysical feedback.
For instance, reionization fronts can halt star formation and accretion in halos with shallow potential wells~\citep{Shapiro:2003gxa}. 
In order to account for this degeneracy we add a turnover to the SFE as~\citet{Park:2018ljd}
\be
f_\star(M_h) \to f_\star(M_h) \times e^{-M_{\rm turn}/M_h}
\ee
which depends on a threshold mass $M_{\rm turn}$, by default set to the atomic-cooling threshold~\citep{Oh:2001ex} but which we will now vary as a free parameter.

\begin{figure}
    \centering
    \includegraphics[width=0.96\linewidth]{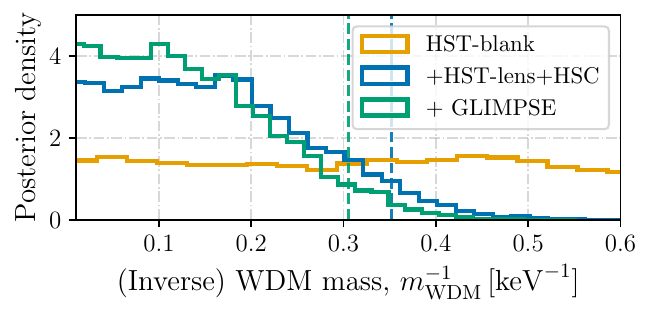}
    \caption{Posterior for the (inverse) WDM mass for different datasets. Adding GLIMPSE (green) tightens the 2$\sigma$ limit (vertical dashed lines), reaching $m_{\rm WDM} > 3.2$ keV, the strongest reionization-era limit to date. 
    }
    \label{fig:corner_WDM_Mturn}
\end{figure}

We perform a WDM search where we fit the UVLFs and bias varying our astrophysical parameters as above, plus $\log_{10} M_{\rm turn}/M_\odot \in [5,9]$, the HMF parameter $a_{\rm ST} \in [0.9,1]$, and the WDM mass (through $m_{\rm WDM}^{-1}$ as is convention, in the $[0,1]$ keV$^{-1}$ range).
Our analysis does not prefer WDM over CDM ($\Delta \chi^2=0.7$), and  
Fig.~\ref{fig:corner_WDM_Mturn} shows the posterior for the WDM mass using different datasets. 
HST-blank alone can only reach very weak constraints on $m_{\rm WDM}$, which improve when adding HST-lens and Subaru/HSC clustering.
Adding GLIMPSE strengthens the constraints to $m_{\rm WDM} > 3.2$ keV at 95\% CL. A comparable limit was obtained in \citet{Liu:2024edl}, though relying on $z\gtrsim 9$ UVLFs and marginalizing over fewer astrophysical parameters (e.g., $f_\star$ and $\sigma_{\rm UV}$ were not allowed to evolve from $z\sim 4-14$).
Our WDM bound is weaker than limits from lower-$z$ probes, which can reach $\approx 6$ keV individually~\citep[as reviewed in][]{Nadler:2026waz} and up to $\approx 10$ keV when combined~\citep{Nadler:2021dft}. 
Fig.~\ref{fig:Pk_result} again contextualizes our limit, where WDM departs from standard CDM within the scales that we probe.

In addition to WDM, we find an upper bound on the turnover mass of $M_{\rm turn} \leq 10^{8.4}\,\Msun$ (95\% CL), given the observed UVLFs keep growing towards the faint end.
This provides an empirical limit on the strength of reionization and other feedback processes regulating galaxy formation at $z\sim 7$.

\section{Discussion and conclusions}

In this {\it Letter} we use new determinations of the UVLF from GLIMPSE, the deepest JWST lensing program to date, to measure the matter power spectrum $P(k)$ at $z\sim 4-8$. 
Marginalizing over astrophysical uncertainties, we measure the amplitude of $P(k)$ to be in agreement with CDM within 0.2 dex at $k=3-9\,\Mpcinv$ and 0.5 dex at $k=9-27\,\Mpcinv$.  
These DM-model-agnostic constraints improve upon past high-$z$ work both in precision and reach, and can be translated into limits for different models.
For the specific case of WDM we find $m_{\rm WDM} > 3.2$ keV at 95\% CL. 
While lower-$z$ observations can reach smaller scales (and consequently better WDM limits), they rely on different systematics and modeling. 
Any departure from CDM has to be confirmed across redshifts to ensure that it is physical, and our analysis extends this test to $z\sim 7$ (or 750 Myr after the Big Bang), providing an independent handle on the nature of DM.

Our analysis relies on an empirical model that connects DM halos to galaxies with free parameters that are varied alongside $P(k)$. 
Given our imperfect knowledge of early galaxy formation, some caveats apply. 
For instance, the role of dust at high $z$ is not entirely settled~\citep{Narayanan18_dust,McKinney:2025htm}. To account for this potential systematic we have repeated our analyses varying the dust calibration in the fit (within the priors $C_{1,\rm dust} \in [1.1,2.1]$ and $\beta_{0,\rm dust} \in [-2.6,-1.6]$,  bracketed by the calibrations of \citealt{Casey:2014cqa} and \citealt{Overzier:2010aa}). 
We have also enhanced the halo-galaxy connection by allowing the pivot mass $M_p$ to vary with $z$ as well.
The results of these tests are shown in Appendix~\ref{app:robustness}; in all cases we find the cosmological constraints nearly identical to the ones presented above. 
On the simulation side, the HMF is not precisely understood at high $z$ beyond CDM, which contributes to our error-bars through parameter marginalization over $a_{\rm ST}$ (and we verify robustness to the window parameter $c$ in Appendix~\ref{app:robustness}). We hope this work encourages further high-$z$ simulations beyond CDM to improve upon these uncertainties. 
Finally, the lens model and cosmic variance both contribute significantly to the faint-end UVLF errors~\citep{Atek:2018nsc}. 
We use the conservative error-bars from \citet{Atek_26_UVLFGLimpse}, which use different models to marginalize over this uncertainty, and note that an independent re-analysis in~\citet{Goolsby_26_UVLFturnover_usingGLimpse} found similar UVLFs. 
To test the impact of the faint-end uncertainties in our analysis (as well as any potential contribution from globular clusters, \citealt{Boylan-Kolchin:2017qaw}), we have re-run our analysis removing the faintest $\MUV > -14$ bins, and find nearly identical $P(k)$ constraints (also in Appendix~\ref{app:robustness}), showing that our results are not driven by the faintest few UVLF datapoints.

Looking ahead, the next decade will bring powerful data to test early structure formation. 
Determinations of the faint end of the UVLF will improve with deeper blank-field observations and multi-cluster lensed programs~\citep[e.g.,][]{Nakane_25_Venus}; whereas measurements from the {\it Nancy Grace Roman} Space Telescope will essentially remove cosmic variance  at the bright end~\citep{Wang:2021oec} and anchor the halo-galaxy connection through clustering~\citep{Munoz:2023cup}. 
Line-intensity maps (both 21-cm and otherwise) are expected to measure the abundance of sources up to higher redshifts and smaller masses than direct methods (see~\citealt{Bernal:2022jap} for a recent review), increasing the reach in $P(k)$ up to $k\sim 100\,\Mpcinv$~\citep{Munoz:2019hjh}. 
In the meantime, improved modeling can ``squeeze'' more information out of current data.
For instance, using the full 2-point correlation function (vs just the galaxy bias) will further break degeneracies and aid in measuring $P(k)$. 
Moreover, an improved understanding of the astrophysics of the $z\gtrsim 9$ UVLF \citep[e.g.,][]{Donnan24_UVLF,Chemerynska_26_GlimpseUVLF} would allow us to use those data to constrain structure formation, as they are highly sensitive to departures from CDM.

To summarize, we have used HST, Subaru and JWST
observations to measure how DM clusters on sub-Galactic scales during the first billion years. We find agreement with the prediction of the standard cold DM model up to wavenumbers $k\sim 27\,\Mpcinv$, and set the strongest reionization-era bounds on warm DM. These results show how early-Universe galaxies can probe the nature of DM in a new range of cosmic times and scales.

\section*{Acknowledgements}

We are grateful to Mike Boylan-Kolchin for discussions. 
J.B.M.~was supported by the National Science Foundation under Grants AST-2307354 and AST-2408637, and NASA through grant JWST-GO-03224. 
Support for US investigators in program \#3293 was provided by NASA through a grant from the Space Telescope Science Institute, which is operated by the Association of Universities for Research in Astronomy, Inc., under NASA contract NAS5-03127. H.A. acknowledges support from CNES, focused on the JWST mission, and the French National Research Agency (ANR) under grant ANR-21-CE31-0838. P.N. acknowledges support from the Gordon and Betty Moore Foundation and the John Templeton Foundation, which fund the Black Hole Initiative (BHI) at Harvard University, where she serves as a PI. P.N. also acknowledges support from the John Templeton Foundation via grant 126613. A.S.L. acknowledges support from Knut and Alice Wallenberg Foundation.

%%%%%%%%%%%%%%%%%%%%%%%%%%%%%%%%%%%%%%%%%%%%%%%%%%
\section*{Data Availability}

The observations used in this paper are publicly available, and we make our codebase public on Github as part of {\tt Zeus21}.

\software{numpy~\citep{numpy}, scipy~\citep{scipy}, matplotlib~\citep{matplotlib},
CLASS~\citep{Blas:2011rf},
Zeus21~\citep{Munoz:2023kkg,Cruz:2024fsv}. }

\section*{AI Usage}

Generative AI tools (Claude Opus v5.5) were used for spellchecks, minor edits, literature search, and code improvements. All AI-generated code and text were reviewed by the authors.



\bibliographystyle{aasjournal}
\bibliography{jwst} % if your bibtex file is called example.bib

@ARTICLE{Jecmen26_betaUV,
       author = {{Jecmen}, Michelle C. and {Chisholm}, John and {Atek}, Hakim and {Kokorev}, Vasily and {Endsley}, Ryan and {Chemerynska}, Iryna and {Furtak}, Lukas J. and {Pan}, Richard and {Fujimoto}, Seiji and {Naidu}, Rohan P. and {Mu{\~n}oz}, Julian B. and {Adamo}, Angela and {Asada}, Yoshihisa and {Basu}, Arghyadeep and {Berg}, Danielle A. and {Blaizot}, Jeremy and {Dessauges-Zavadsky}, Miroslava and {Giovinazzo}, Emma and {Hsiao}, Tiger Yu-Yang and {Katz}, Harley and {Korber}, Damien and {McKinney}, Jed and {McQuinn}, Kristen. B.~W. and {Oesch}, Pascal A. and {Lopez}, A. Saldana- and {Schaerer}, Daniel},
        title = "{A GLIMPSE into the UV Continuum Slopes of the Faintest Galaxies in the Epoch of Reionization}",
      journal = {\apj},
         year = 2026,
        month = apr,
       volume = {1001},
       number = {2},
          eid = {234},
        pages = {234},
          doi = {10.3847/1538-4357/ae5246},
archivePrefix = {arXiv},
       eprint = {2601.19995},
 primaryClass = {astro-ph.GA},
       adsurl = {https://ui.adsabs.harvard.edu/abs/2026ApJ..1001..234J}
}

@ARTICLE{Natarajan+2026,
       author = {{Natarajan}, Priyamvada and {Chiang}, Barry T. and {Dutra}, Isaque},
        title = "{New Cold Dark Matter Crisis Revealed by Multiscale Cluster Lensing}",
      journal = {\apjl},
         year = 2026,
        month = apr,
       volume = {1001},
       number = {1},
          eid = {L12},
        pages = {L12},
          doi = {10.3847/2041-8213/ae53ea},
archivePrefix = {arXiv},
       eprint = {2601.07909},
 primaryClass = {astro-ph.CO},
       adsurl = {https://ui.adsabs.harvard.edu/abs/2026ApJ..1001L..12N}
}

@article{Boylan-Kolchin:2017qaw,
    author = "Boylan-Kolchin, Michael",
    title = "{The Little Engines That Could? Globular Clusters Contribute Significantly to Reionization-era Star Formation}",
    eprint = "1711.00009",
    archivePrefix = "arXiv",
    primaryClass = "astro-ph.GA",
    doi = "10.1093/mnras/sty1490",
    journal = "Mon. Not. Roy. Astron. Soc.",
    volume = "479",
    number = "1",
    pages = "332--340",
    year = "2018"
}

@article{Dekker:2024nkb,
    author = "Dekker, Ariane and Kravtsov, Andrey",
    title = "{Constraints on blue and red tilted primordial power spectra using dwarf galaxy properties}",
    eprint = "2407.04198",
    archivePrefix = "arXiv",
    primaryClass = "astro-ph.CO",
    doi = "10.1103/PhysRevD.111.063516",
    journal = "Phys. Rev. D",
    volume = "111",
    number = "6",
    pages = "063516",
    year = "2025"
}

@article{Benson:2026kdp,
    author = "Benson, Andrew J. and Nadler, Ethan O. and Du, Xiaolong and Gluscevic, Vera",
    title = "{A Unified Halo Mass Function Across Dark Matter Models from High-Resolution Multi-Scale Simulations}",
    eprint = "2606.12137",
    archivePrefix = "arXiv",
    primaryClass = "astro-ph.CO",
    month = "6",
    year = "2026"
}

@article{Vogelsberger:2015gpr,
    author = "Vogelsberger, Mark and Zavala, Jesus and Cyr-Racine, Francis-Yan and Pfrommer, Christoph and Bringmann, Torsten and Sigurdson, Kris",
    title = "{ETHOS {\textendash} an effective theory of structure formation: dark matter physics as a possible explanation of the small-scale CDM problems}",
    eprint = "1512.05349",
    archivePrefix = "arXiv",
    primaryClass = "astro-ph.CO",
    doi = "10.1093/mnras/stw1076",
    journal = "Mon. Not. Roy. Astron. Soc.",
    volume = "460",
    number = "2",
    pages = "1399--1416",
    year = "2016"
}

@ARTICLE{Nakane_25_Venus,
       author = {{Nakane}, Minami and {Kokorev}, Vasily and {Fujimoto}, Seiji and {Ouchi}, Masami and {McLeod}, Derek J. and {Golubchik}, Miriam and {Oguri}, Masamune and {Zitrin}, Adi and {Bondestam}, Cecilia and {Donnan}, Callum T. and {Brammer}, Gabriel and {Finkelstein}, Steven L. and {Willott}, Chris and {Rihtarsic}, Gregor and {Desprez}, Guillaume and {Adamo}, Angela and {Vanzella}, Eros and {Brada{\v{c}}}, Maru{\v{s}}a and {Messa}, Matteo and {Yanagisawa}, Hiroto and {Sun}, Fengwu and {Ferguson}, Henry C. and {Lucas}, Ray A. and {Coe}, Dan and {Richard}, Johan and {Abdurro'uf} and {Akins}, Hollis B. and {Allingham}, Joseph F.~V. and {Amor{\'\i}n}, Ricardo O. and {Asada}, Yoshihisa and {Atek}, Hakim and {Bezanson}, Rachel and {Bradley}, Larry D. and {Chisholm}, John and {Conselice}, Christopher J. and {Dayal}, Pratika and {Dessauges-Zavadsky}, Miroslava and {Diego}, Jose M. and {Faisst}, Andreas L. and {Farley}, Gavin and {Fei}, Qinyue and {Frye}, Brenda L. and {Fudamoto}, Yoshinobu and {Furtak}, Lukas J. and {Harikane}, Yuichi and {Hsiao}, Tiger Yu-Yang and {Jim{\'e}nez-Teja}, Yolanda and {Kartaltepe}, Jeyhan S. and {Kiyota}, Tomokazu and {Koekemoer}, Anton M. and {Lagos}, Claudia del P. and {Magdis}, Georgios E. and {Meena}, Ashish Kumar and {Mowla}, Lamiya and {Noirot}, Ga{\"e}l and {Oesch}, Pascal A. and {Ono}, Yoshiaki and {Ortiz}, III, Rafael and {Pan}, Richard and {Papovich}, Casey and {Pierel}, Justin D. and {Ricotti}, Massimo and {Robbins}, Luke and {Schaerer}, Daniel and {Schneider}, Raffaella and {Treu}, Tommaso and {Valentino}, Francesco and {Windhorst}, Rogier A. and {Bauer}, Franz E. and {Bromm}, Volker and {Egami}, Eiichi and {Gonz{\'a}lez-Otero}, Mauro and {Kohno}, Kotaro and {Labbe}, Ivo and {Matthee}, Jorryt and {Mun}, Marcie and {Naidu}, Rohan P. and {Tripodi}, Roberta},
        title = "{VENUS: A Strongly Lensed Clumpy Galaxy at $z\sim11-12$ behind the Galaxy Cluster MACS J0257.1-2325}",
      journal = {arXiv e-prints},
         year = 2025,
        month = nov,
          eid = {arXiv:2511.14483},
        pages = {arXiv:2511.14483},
          doi = {10.48550/arXiv.2511.14483},
archivePrefix = {arXiv},
       eprint = {2511.14483},
 primaryClass = {astro-ph.GA},
       adsurl = {https://ui.adsabs.harvard.edu/abs/2025arXiv251114483N}
}

@ARTICLE{Chiang+2026,
       author = {{Chiang}, Barry T. and {Dutra}, Isaque and {Natarajan}, Priyamvada},
        title = "{Constraining the Nature of Dark Matter from Tidal Radii of Cluster Galaxy Subhalos}",
      journal = {\apj},
         year = 2026,
        month = jan,
       volume = {997},
       number = {1},
          eid = {106},
        pages = {106},
          doi = {10.3847/1538-4357/ae23c1},
archivePrefix = {arXiv},
       eprint = {2511.14726},
 primaryClass = {astro-ph.CO},
       adsurl = {https://ui.adsabs.harvard.edu/abs/2026ApJ...997..106C}
}

@MISC{scipy,
	author =    {Eric Jones and Travis Oliphant and Pearu Peterson and others},
	title =     {{SciPy}: Open source scientific tools for {Python}},
	year =      {2001},
	url = "http://www.scipy.org/",
	note = {[Online; accessed <today>]}
}

@ARTICLE{matplotlib,
	author = {{Hunter}, John D.},
	title = "{Matplotlib: A 2D Graphics Environment}",
	journal = {Computing in Science and Engineering},
	year = "2007",
	month = "May",
	volume = {9},
	number = {3},
	pages = {90-95},
	doi = {10.1109/MCSE.2007.55},
	adsurl = {https://ui.adsabs.harvard.edu/abs/2007CSE.....9...90H}
}

@ARTICLE{numpy,
	author = {{van der Walt}, St{\'e}fan and {Colbert}, S. Chris and
	{Varoquaux}, Ga{\"e}l},
	title = "{The NumPy Array: A Structure for Efficient Numerical Computation}",
	journal = {Computing in Science and Engineering},
	year = "2011",
	month = "Mar",
	volume = {13},
	number = {2},
	pages = {22-30},
	doi = {10.1109/MCSE.2011.37},
	archivePrefix = {arXiv},
	eprint = {1102.1523},
	primaryClass = {cs.MS},
	adsurl = {https://ui.adsabs.harvard.edu/abs/2011CSE....13b..22V}
}

@article{Blas:2011rf,
	author         = "Blas, Diego and Lesgourgues, Julien and Tram, Thomas",
	title          = "{The Cosmic Linear Anisotropy Solving System (CLASS) II:
	Approximation schemes}",
	journal        = "JCAP",
	volume         = "1107",
	year           = "2011",
	pages          = "034",
	doi            = "10.1088/1475-7516/2011/07/034",
	eprint         = "1104.2933",
	archivePrefix  = "arXiv",
	
	reportNumber   = "CERN-PH-TH-2011-082, LAPTH-010-11",
	SLACcitation   = "%%CITATION = ARXIV:1104.2933;%%"
}

@ARTICLE{FinkelsteinBagley22_UVLF,
       author = {{Finkelstein}, Steven L. and {Bagley}, Micaela B.},
        title = "{On the Coevolution of the AGN and Star-forming Galaxy Ultraviolet Luminosity Functions at 3 < z < 9}",
      journal = {\apj},
         year = 2022,
        month = oct,
       volume = {938},
       number = {1},
          eid = {25},
        pages = {25},
          doi = {10.3847/1538-4357/ac89eb},
archivePrefix = {arXiv},
       eprint = {2207.02233},
 primaryClass = {astro-ph.GA},
       adsurl = {https://ui.adsabs.harvard.edu/abs/2022ApJ...938...25F}
}

@ARTICLE{OkeGunnAB,
       author = {{Oke}, J.~B. and {Gunn}, J.~E.},
        title = "{Secondary standard stars for absolute spectrophotometry.}",
      journal = {\apj},
         year = 1983,
        month = mar,
       volume = {266},
        pages = {713-717},
          doi = {10.1086/160817},
       adsurl = {https://ui.adsabs.harvard.edu/abs/1983ApJ...266..713O}
}

@article{Bouwens:2013hxa,
    author = "Bouwens, R.J. and others",
    title = "{UV-Continuum Slopes of >4000 $z \sim$ 4-8 Galaxies from the HUDF/XDF, HUDF09, ERS, CANDELS-South, and CANDELS-North Fields}",
    eprint = "1306.2950",
    archivePrefix = "arXiv",
    primaryClass = "astro-ph.CO",
    doi = "10.1088/0004-637X/793/2/115",
    journal = "Astrophys. J.",
    volume = "793",
    number = "2",
    pages = "115",
    year = "2014"
}

@ARTICLE{Bouwens16xiion,
       author = {{Bouwens}, Rychard J. and {Aravena}, Manuel and {Decarli}, Roberto and {Walter}, Fabian and {da Cunha}, Elisabete and {Labb{\'e}}, Ivo and {Bauer}, Franz E. and {Bertoldi}, Frank and {Carilli}, Chris and {Chapman}, Scott and {Daddi}, Emanuele and {Hodge}, Jacqueline and {Ivison}, Rob J. and {Karim}, Alex and {Le Fevre}, Olivier and {Magnelli}, Benjamin and {Ota}, Kazuaki and {Riechers}, Dominik and {Smail}, Ian R. and {van der Werf}, Paul and {Weiss}, Axel and {Cox}, Pierre and {Elbaz}, David and {Gonzalez-Lopez}, Jorge and {Infante}, Leopoldo and {Oesch}, Pascal and {Wagg}, Jeff and {Wilkins}, Steve},
        title = "{ALMA Spectroscopic Survey in the Hubble Ultra Deep Field: The Infrared Excess of UV-Selected z = 2-10 Galaxies as a Function of UV-Continuum Slope and Stellar Mass}",
      journal = {\apj},
         year = 2016,
        month = dec,
       volume = {833},
       number = {1},
          eid = {72},
        pages = {72},
          doi = {10.3847/1538-4357/833/1/72},
archivePrefix = {arXiv},
       eprint = {1606.05280},
 primaryClass = {astro-ph.GA},
       adsurl = {https://ui.adsabs.harvard.edu/abs/2016ApJ...833...72B}
}

@ARTICLE{Donnan24_UVLF,
       author = {{Donnan}, C.~T. and {McLure}, R.~J. and {Dunlop}, J.~S. and {McLeod}, D.~J. and {Magee}, D. and {Arellano-C{\'o}rdova}, K.~Z. and {Barrufet}, L. and {Begley}, R. and {Bowler}, R.~A.~A. and {Carnall}, A.~C. and {Cullen}, F. and {Ellis}, R.~S. and {Fontana}, A. and {Illingworth}, G.~D. and {Grogin}, N.~A. and {Hamadouche}, M.~L. and {Koekemoer}, A.~M. and {Liu}, F.-Y. and {Mason}, C. and {Santini}, P. and {Stanton}, T.~M.},
        title = "{JWST PRIMER: a new multifield determination of the evolving galaxy UV luminosity function at redshifts z ≃ 9 - 15}",
      journal = {\mnras},
         year = 2024,
        month = sep,
       volume = {533},
       number = {3},
        pages = {3222-3237},
          doi = {10.1093/mnras/stae2037},
archivePrefix = {arXiv},
       eprint = {2403.03171},
 primaryClass = {astro-ph.GA},
       adsurl = {https://ui.adsabs.harvard.edu/abs/2024MNRAS.533.3222D}
}

@ARTICLE{Ciesla_24_burstiness,
       author = {{Ciesla}, L. and {Elbaz}, D. and {Ilbert}, O. and {Buat}, V. and {Magnelli}, B. and {Narayanan}, D. and {Daddi}, E. and {G{\'o}mez-Guijarro}, C. and {Arango-Toro}, R.},
        title = "{Identification of a transition from stochastic to secular star formation around z = 9 with JWST}",
      journal = {\aap},
         year = 2024,
        month = jun,
       volume = {686},
          eid = {A128},
        pages = {A128},
          doi = {10.1051/0004-6361/202348091},
archivePrefix = {arXiv},
       eprint = {2309.15720},
 primaryClass = {astro-ph.GA},
       adsurl = {https://ui.adsabs.harvard.edu/abs/2024A&A...686A.128C}
}

@ARTICLE{Sun25_burstiness,
       author = {{Sun}, Guochao and {Mu{\~n}oz}, Julian B. and {Mirocha}, Jordan and {Faucher-Gigu{\`e}re}, Claude-Andr{\'e}},
        title = "{Constraining bursty star formation histories with galaxy UV and H{\ensuremath{\alpha}} luminosity functions and clustering}",
      journal = {\jcap},
         year = 2025,
        month = apr,
       volume = {2025},
       number = {4},
          eid = {034},
        pages = {034},
          doi = {10.1088/1475-7516/2025/04/034},
archivePrefix = {arXiv},
       eprint = {2410.21409},
 primaryClass = {astro-ph.GA},
       adsurl = {https://ui.adsabs.harvard.edu/abs/2025JCAP...04..034S}
}

@article{Munoz:2026gxp,
    author = "Mu{\~n}oz, Julian B. and Chisholm, John and Sun, Guochao and Samuel, Jenna and Mirocha, Jordan and Bregou, Emily and Venditti, Alessandra and Qezlou, Mahdi and Simmonds, Charlotte and Endsley, Ryan",
    title = "{Relatively fast and reasonably furious: evidence for increased burstiness in smaller haloes at cosmic dawn}",
    eprint = "2601.07912",
    archivePrefix = "arXiv",
    primaryClass = "astro-ph.GA",
    doi = "10.1093/mnras/stag415",
    journal = "Mon. Not. Roy. Astron. Soc.",
    volume = "547",
    number = "4",
    pages = "stag415",
    year = "2026"
}

@ARTICLE{Endsley25_burstiness,
       author = {{Endsley}, Ryan and {Chisholm}, John and {Stark}, Daniel P. and {Topping}, Michael W. and {Whitler}, Lily},
        title = "{The Burstiness of Star Formation at z {\ensuremath{\sim}} 6: A Huge Diversity in the Recent Star Formation Histories of Very UV-faint Galaxies}",
      journal = {\apj},
         year = 2025,
        month = jul,
       volume = {987},
       number = {2},
          eid = {189},
        pages = {189},
          doi = {10.3847/1538-4357/addc74},
archivePrefix = {arXiv},
       eprint = {2410.01905},
 primaryClass = {astro-ph.GA},
       adsurl = {https://ui.adsabs.harvard.edu/abs/2025ApJ...987..189E}
}

@ARTICLE{Bouwens21,
       author = {{Bouwens}, R.~J. and {Oesch}, P.~A. and {Stefanon}, M. and {Illingworth}, G. and {Labb{\'e}}, I. and {Reddy}, N. and {Atek}, H. and {Montes}, M. and {Naidu}, R. and {Nanayakkara}, T. and {Nelson}, E. and {Wilkins}, S.},
        title = "{New Determinations of the UV Luminosity Functions from z   9 to 2 Show a Remarkable Consistency with Halo Growth and a Constant Star Formation Efficiency}",
      journal = {\aj},
         year = 2021,
        month = aug,
       volume = {162},
       number = {2},
          eid = {47},
        pages = {47},
          doi = {10.3847/1538-3881/abf83e},
archivePrefix = {arXiv},
       eprint = {2102.07775},
 primaryClass = {astro-ph.GA},
       adsurl = {https://ui.adsabs.harvard.edu/abs/2021AJ....162...47B}
}

@article{Liu:2024edl,
    author = "Liu, Bin and Shan, Huanyuan and Zhang, Jiajun",
    title = "{New Galaxy UV Luminosity Constraints on Warm Dark Matter from JWST}",
    eprint = "2404.13596",
    archivePrefix = "arXiv",
    primaryClass = "astro-ph.CO",
    doi = "10.3847/1538-4357/ad4ed8",
    journal = "Astrophys. J.",
    volume = "968",
    number = "2",
    pages = "79",
    year = "2024"
}

@article{Nadler:2021dft,
    author = "Nadler, Ethan O. and Birrer, Simon and Gilman, Daniel and Wechsler, Risa H. and Du, Xiaolong and Benson, Andrew and Nierenberg, Anna M. and Treu, Tommaso",
    title = "{Dark Matter Constraints from a Unified Analysis of Strong Gravitational Lenses and Milky Way Satellite Galaxies}",
    eprint = "2101.07810",
    archivePrefix = "arXiv",
    primaryClass = "astro-ph.CO",
    doi = "10.3847/1538-4357/abf9a3",
    journal = "Astrophys. J.",
    volume = "917",
    number = "1",
    pages = "7",
    year = "2021"
}

@ARTICLE{Planck:2018vyg,
    author = "Aghanim, N. and others",
    collaboration = "Planck",
    title = "{Planck 2018 results. VI. Cosmological parameters}",
    eprint = "1807.06209",
    archivePrefix = "arXiv",
    primaryClass = "astro-ph.CO",
    doi = "10.1051/0004-6361/201833910",
    journal = "Astron. Astrophys.",
    volume = "641",
    pages = "A6",
    year = "2020",
    note = "[Erratum: Astron.Astrophys. 652, C4 (2021)]"
}

@article{Berlind:2001xk,
    author = "Berlind, Andreas A. and Weinberg, David H.",
    title = "{The Halo occupation distribution: Towards an empirical determination of the relation between galaxies and mass}",
    eprint = "astro-ph/0109001",
    archivePrefix = "arXiv",
    doi = "10.1086/341469",
    journal = "Astrophys. J.",
    volume = "575",
    pages = "587--616",
    year = "2002"
}

@article{Munoz:2023cup,
    author = "Mu\~noz, Julian B. and Mirocha, Jordan and Furlanetto, Steven and Sabti, Nashwan",
    title = "{Breaking degeneracies in the first galaxies with clustering}",
    eprint = "2306.09403",
    archivePrefix = "arXiv",
    primaryClass = "astro-ph.CO",
    doi = "10.1093/mnrasl/slad115",
    journal = "Mon. Not. Roy. Astron. Soc.",
    volume = "526",
    number = "1",
    pages = "L47--L55",
    year = "2023"
}

@article{Munoz:2023kkg,
    author = "Mu{\~n}oz, Julian B.",
    title = "{An effective model for the cosmic-dawn 21-cm signal}",
    eprint = "2302.08506",
    archivePrefix = "arXiv",
    primaryClass = "astro-ph.CO",
    doi = "10.1093/mnras/stad1512",
    journal = "Mon. Not. Roy. Astron. Soc.",
    volume = "523",
    number = "2",
    pages = "2587--2607",
    year = "2023"
}

@article{Madau:2014bja,
    author = "Madau, Piero and Dickinson, Mark",
    title = "{Cosmic Star Formation History}",
    eprint = "1403.0007",
    archivePrefix = "arXiv",
    primaryClass = "astro-ph.CO",
    doi = "10.1146/annurev-astro-081811-125615",
    journal = "Ann. Rev. Astron. Astrophys.",
    volume = "52",
    pages = "415--486",
    year = "2014"
}

@article{Marsh:2013ywa,
    author = "Marsh, David J. E. and Silk, Joe",
    title = "{A Model For Halo Formation With Axion Mixed Dark Matter}",
    eprint = "1307.1705",
    archivePrefix = "arXiv",
    primaryClass = "astro-ph.CO",
    doi = "10.1093/mnras/stt2079",
    journal = "Mon. Not. Roy. Astron. Soc.",
    volume = "437",
    number = "3",
    pages = "2652--2663",
    year = "2014"
}

@article{Schneider:2011yu,
    author = "Schneider, Aurel and Smith, Robert E. and Maccio, Andrea V. and Moore, Ben",
    title = "{Nonlinear Evolution of Cosmological Structures in Warm Dark Matter Models}",
    eprint = "1112.0330",
    archivePrefix = "arXiv",
    primaryClass = "astro-ph.CO",
    doi = "10.1111/j.1365-2966.2012.21252.x",
    journal = "Mon. Not. Roy. Astron. Soc.",
    volume = "424",
    pages = "684",
    year = "2012"
}

@article{Shen:2023cva,
       author = {{Shen}, Xuejian and {Vogelsberger}, Mark and {Boylan-Kolchin}, Michael and {Tacchella}, Sandro and {Kannan}, Rahul},
        title = "{The impact of UV variability on the abundance of bright galaxies at z {\ensuremath{\geq}} 9}",
      journal = {\mnras},
         year = 2023,
        month = nov,
       volume = {525},
       number = {3},
        pages = {3254-3261},
          doi = {10.1093/mnras/stad2508},
archivePrefix = {arXiv},
       eprint = {2305.05679},
 primaryClass = {astro-ph.GA},
       adsurl = {https://ui.adsabs.harvard.edu/abs/2023MNRAS.525.3254S}
}

@ARTICLE{Mirocha_UVLFs2023,
       author = {{Mirocha}, Jordan and {Furlanetto}, Steven R.},
        title = "{Balancing the efficiency and stochasticity of star formation with dust extinction in z {\ensuremath{\gtrsim}} 10 galaxies observed by JWST}",
      journal = {Mon. Notices Royal Astron. Soc.},
         year = 2023,
        month = feb,
       volume = {519},
       number = {1},
        pages = {843-853},
          doi = {10.1093/mnras/stac3578},
archivePrefix = {arXiv},
       eprint = {2208.12826},
 primaryClass = {astro-ph.GA},
       adsurl = {https://ui.adsabs.harvard.edu/abs/2023MNRAS.519..843M}
}

@article{Shapiro:2003gxa,
    author = "Shapiro, Paul R. and Iliev, Ilian T. and Raga, Alejandro C.",
    title = "{Photoevaporation of cosmological minihalos during reionization}",
    eprint = "astro-ph/0307266",
    archivePrefix = "arXiv",
    doi = "10.1111/j.1365-2966.2004.07364.x",
    journal = "Mon. Not. Roy. Astron. Soc.",
    volume = "348",
    pages = "753",
    year = "2004"
}

@article{Gilman:2021gkj,
    author = "Gilman, Daniel and Benson, Andrew and Bovy, Jo and Birrer, Simon and Treu, Tommaso and Nierenberg, Anna",
    title = "{The primordial matter power spectrum on sub-galactic scales}",
    eprint = "2112.03293",
    archivePrefix = "arXiv",
    primaryClass = "astro-ph.CO",
    doi = "10.1093/mnras/stac670",
    journal = "Mon. Not. Roy. Astron. Soc.",
    volume = "512",
    number = "3",
    pages = "3163--3188",
    year = "2022"
}

@article{deKruijf:2024voc,
    author = "de Kruijf, Jessie and Vanzan, Eleonora and Boddy, Kimberly K. and Raccanelli, Alvise and Bartolo, Nicola",
    title = "{Searching for blue-tilted power spectra in the dark ages}",
    eprint = "2408.04991",
    archivePrefix = "arXiv",
    primaryClass = "astro-ph.CO",
    reportNumber = "UTWI-26-2024",
    doi = "10.1103/PhysRevD.111.063507",
    journal = "Phys. Rev. D",
    volume = "111",
    number = "6",
    pages = "063507",
    year = "2025"
}

@article{Wang:2021oec,
    author = "Wang, Yun and others",
    title = "{The High Latitude Spectroscopic Survey on the Nancy Grace Roman Space Telescope}",
    eprint = "2110.01829",
    archivePrefix = "arXiv",
    primaryClass = "astro-ph.CO",
    doi = "10.3847/1538-4357/ac4973",
    journal = "Astrophys. J.",
    volume = "928",
    number = "1",
    pages = "1",
    year = "2022"
}

@article{Bernal:2022jap,
    author = "Bernal, Jos{\'e} Luis and Kovetz, Ely D.",
    title = "{Line-intensity mapping: theory review with a focus on star-formation lines}",
    eprint = "2206.15377",
    archivePrefix = "arXiv",
    primaryClass = "astro-ph.CO",
    doi = "10.1007/s00159-022-00143-0",
    journal = "Astron. Astrophys. Rev.",
    volume = "30",
    number = "1",
    pages = "5",
    year = "2022"
}

@article{Atek:2018nsc,
    author = "Atek, Hakim and Richard, Johan and Kneib, Jean-Paul and Schaerer, Daniel",
    title = "{The Extreme Faint End of the UV Luminosity Function at $z\sim6$ Through Gravitational Telescopes: a comprehensive assessment of strong lensing uncertainties}",
    eprint = "1803.09747",
    archivePrefix = "arXiv",
    primaryClass = "astro-ph.GA",
    doi = "10.1093/mnras/sty1820",
    journal = "Mon. Not. Roy. Astron. Soc.",
    volume = "479",
    number = "4",
    pages = "5184--5195",
    year = "2018"
}

@article{Sabti:2021unj,
    author = "Sabti, Nashwan and Mu\~noz, Julian B. and Blas, Diego",
    title = "{New Roads to the Small-scale Universe: Measurements of the Clustering of Matter with the High-redshift UV Galaxy Luminosity Function}",
    eprint = "2110.13161",
    archivePrefix = "arXiv",
    primaryClass = "astro-ph.CO",
    reportNumber = "KCL-2021-75",
    doi = "10.3847/2041-8213/ac5e9c",
    journal = "Astrophys. J. Lett.",
    volume = "928",
    number = "2",
    pages = "L20",
    year = "2022"
}

@article{Tacchella:2018qny,
    author = "Tacchella, Sandro and Bose, Sownak and Conroy, Charlie and Eisenstein, Daniel J. and Johnson, Benjamin D.",
    title = "{A Redshift-independent Efficiency Model: Star Formation and Stellar Masses in Dark Matter Halos at $z \geq 4$}",
    eprint = "1806.03299",
    archivePrefix = "arXiv",
    primaryClass = "astro-ph.GA",
    doi = "10.3847/1538-4357/aae8e0",
    journal = "Astrophys. J.",
    volume = "868",
    number = "2",
    pages = "92",
    year = "2018"
}

@article{Meurer:1999jj,
    author = "Meurer, Gerhardt R. and Heckman, Timothy M. and Calzetti, Daniela",
    title = "{Dust absorption and the ultraviolet luminosity density at z\textasciitilde\ 3 as calibrated by local starburst galaxies}",
    eprint = "astro-ph/9903054",
    archivePrefix = "arXiv",
    doi = "10.1086/307523",
    journal = "Astrophys. J.",
    volume = "521",
    pages = "64",
    year = "1999"
}

@article{Bouwens:2014fua,
    author = "Bouwens, R.J. and others",
    title = "{UV Luminosity Functions at redshifts $z \sim$4 to $z \sim$10: 10000 Galaxies from HST Legacy Fields}",
    eprint = "1403.4295",
    archivePrefix = "arXiv",
    primaryClass = "astro-ph.CO",
    doi = "10.1088/0004-637X/803/1/34",
    journal = "Astrophys. J.",
    volume = "803",
    number = "1",
    pages = "34",
    year = "2015"
}

@article{Mason:2015cna,
    author = "Mason, Charlotte and Trenti, Michele and Treu, Tommaso",
    title = "{The Galaxy UV Luminosity Function Before the Epoch of Reionization}",
    eprint = "1508.01204",
    archivePrefix = "arXiv",
    primaryClass = "astro-ph.GA",
    doi = "10.1088/0004-637X/813/1/21",
    journal = "Astrophys. J.",
    volume = "813",
    number = "1",
    pages = "21",
    year = "2015"
}

@article{Cruz:2024fsv,
    author = "Cruz, Hector Afonso G. and Munoz, Julian B. and Sabti, Nashwan and Kamionkowski, Marc",
    title = "{Effective model for the 21-cm signal with population III stars}",
    eprint = "2407.18294",
    archivePrefix = "arXiv",
    primaryClass = "astro-ph.CO",
    doi = "10.1103/PhysRevD.111.083503",
    journal = "Phys. Rev. D",
    volume = "111",
    number = "8",
    pages = "083503",
    year = "2025"
}

@article{Oh:2001ex,
    author = "Oh, S.Peng and Haiman, Zoltan",
    title = "{Second-generation objects in the universe: radiative cooling and collapse of halos with virial temperatures above 10^4 kelvin}",
    eprint = "astro-ph/0108071",
    archivePrefix = "arXiv",
    doi = "10.1086/339393",
    journal = "Astrophys. J.",
    volume = "569",
    pages = "558",
    year = "2002"
}

@article{Oesch_2018,
   title={The Dearth of z ∼ 10 Galaxies in All HST Legacy Fields—The Rapid Evolution of the Galaxy Population in the First 500 Myr},
   volume={855},
   ISSN={1538-4357},
   url={http://dx.doi.org/10.3847/1538-4357/aab03f},
   DOI={10.3847/1538-4357/aab03f},
   number={2},
   journal={The Astrophysical Journal},
   publisher={American Astronomical Society},
   author={Oesch, P. A. and Bouwens, R. J. and Illingworth, G. D. and Labb\'{e}, I. and Stefanon, M.},
   year={2018},
   month={Mar},
   pages={105}
}

@article{Wechsler:2018pic,
    author = "Wechsler, Risa H. and Tinker, Jeremy L.",
    title = "{The Connection between Galaxies and their Dark Matter Halos}",
    eprint = "1804.03097",
    archivePrefix = "arXiv",
    primaryClass = "astro-ph.GA",
    doi = "10.1146/annurev-astro-081817-051756",
    journal = "Ann. Rev. Astron. Astrophys.",
    volume = "56",
    pages = "435--487",
    year = "2018"
}

@article{Angulo:2013sza,
    author = "Angulo, Raul E. and Hahn, Oliver and Abel, Tom",
    title = "{The Warm DM halo mass function below the cut-off scale}",
    eprint = "1304.2406",
    archivePrefix = "arXiv",
    primaryClass = "astro-ph.CO",
    doi = "10.1093/mnras/stt1246",
    journal = "Mon. Not. Roy. Astron. Soc.",
    volume = "434",
    pages = "3337",
    year = "2013"
}

@article{Tinker:2008ff,
    author = "Tinker, Jeremy L. and Kravtsov, Andrey V. and Klypin, Anatoly and Abazajian, Kevork and Warren, Michael S. and Yepes, Gustavo and Gottlober, Stefan and Holz, Daniel E.",
    title = "{Toward a halo mass function for precision cosmology: The Limits of universality}",
    eprint = "0803.2706",
    archivePrefix = "arXiv",
    primaryClass = "astro-ph",
    doi = "10.1086/591439",
    journal = "Astrophys. J.",
    volume = "688",
    pages = "709--728",
    year = "2008"
}

@article{Corasaniti:2016epp,
    author = "Corasaniti, P.S. and Agarwal, S. and Marsh, D.J.E. and Das, S.",
    title = "{Constraints on dark matter scenarios from measurements of the galaxy luminosity function at high redshifts}",
    eprint = "1611.05892",
    archivePrefix = "arXiv",
    primaryClass = "astro-ph.CO",
    doi = "10.1103/PhysRevD.95.083512",
    journal = "Phys. Rev. D",
    volume = "95",
    number = "8",
    pages = "083512",
    year = "2017"
}

@article{Menci:2017nsr,
    author = "Menci, Nicola and Merle, Alexander and Totzauer, Maximilian and Schneider, Aurel and Grazian, Andrea and Castellano, Marco and Sanchez, Norma G.",
    title = "{Fundamental physics with the Hubble Frontier Fields: constraining Dark Matter models with the abundance of extremely faint and distant galaxies}",
    eprint = "1701.01339",
    archivePrefix = "arXiv",
    primaryClass = "astro-ph.CO",
    doi = "10.3847/1538-4357/836/1/61",
    journal = "Astrophys. J.",
    volume = "836",
    number = "1",
    pages = "61",
    year = "2017"
}

@article{DES:2021wwk,
    author = "Abbott, T. M. C. and others",
    collaboration = "DES",
    title = "{Dark Energy Survey Year 3 results: Cosmological constraints from galaxy clustering and weak lensing}",
    eprint = "2105.13549",
    archivePrefix = "arXiv",
    primaryClass = "astro-ph.CO",
    reportNumber = "FERMILAB-PUB-21-221-AE, DES-2020-0617",
    doi = "10.1103/PhysRevD.105.023520",
    journal = "Phys. Rev. D",
    volume = "105",
    number = "2",
    pages = "023520",
    year = "2022"
}

@article{Sabti:2021xvh,
    author = "Sabti, Nashwan and Mu\~noz, Julian B. and Blas, Diego",
    title = "{Galaxy luminosity function pipeline for cosmology and astrophysics}",
    eprint = "2110.13168",
    archivePrefix = "arXiv",
    primaryClass = "astro-ph.CO",
    reportNumber = "KCL-2021-74",
    doi = "10.1103/PhysRevD.105.043518",
    journal = "Phys. Rev. D",
    volume = "105",
    number = "4",
    pages = "043518",
    year = "2022"
}

@article{Parashari:2023cui,
    author = "Parashari, Priyank and Laha, Ranjan",
    title = "{Primordial power spectrum in light of JWST observations of high redshift galaxies}",
    eprint = "2305.00999",
    archivePrefix = "arXiv",
    primaryClass = "astro-ph.CO",
    doi = "10.1093/mnrasl/slad107",
    journal = "Mon. Not. Roy. Astron. Soc.",
    volume = "526",
    number = "1",
    pages = "L63--L69",
    year = "2023"
}

@article{Sabti:2020ser,
    author = "Sabti, Nashwan and Mu\~noz, Julian B. and Blas, Diego",
    title = "{First Constraints on Small-Scale Non-Gaussianity from UV Galaxy Luminosity Functions}",
    eprint = "2009.01245",
    archivePrefix = "arXiv",
    primaryClass = "astro-ph.CO",
    reportNumber = "KCL-2020-41",
    doi = "10.1088/1475-7516/2021/01/010",
    journal = "JCAP",
    volume = "01",
    pages = "010",
    year = "2021"
}

@article{Rudakovskyi:2021jyf,
    author = "Rudakovskyi, Anton and Mesinger, Andrei and Savchenko, Denys and Gillet, Nicolas",
    title = "{Constraints on warm dark matter from UV luminosity functions of high-z galaxies with Bayesian model comparison}",
    eprint = "2104.04481",
    archivePrefix = "arXiv",
    primaryClass = "astro-ph.CO",
    doi = "10.1093/mnras/stab2333",
    journal = "Mon. Not. Roy. Astron. Soc.",
    volume = "507",
    number = "2",
    pages = "3046--3056",
    year = "2021"
}

@article{Bowler:2014vra,
    author = "Bowler, R. A. A. and others",
    title = "{The galaxy luminosity function at z {\ensuremath{\simeq}} 6 and evidence for rapid evolution in the bright end from z {\ensuremath{\simeq}} 7 to 5}",
    eprint = "1411.2976",
    archivePrefix = "arXiv",
    primaryClass = "astro-ph.GA",
    doi = "10.1093/mnras/stv1403",
    journal = "Mon. Not. Roy. Astron. Soc.",
    volume = "452",
    number = "2",
    pages = "1817--1840",
    year = "2015"
}

@article{Ono:2017wjz,
    author = "Ono, Yoshiaki and others",
    title = "{Great Optically Luminous Dropout Research Using Subaru HSC (GOLDRUSH). I. UV Luminosity Functions at $z \sim 4-7$ Derived with the Half-Million Dropouts on the 100 deg$^2$ Sky}",
    eprint = "1704.06004",
    archivePrefix = "arXiv",
    primaryClass = "astro-ph.GA",
    doi = "10.1093/pasj/psx103",
    journal = "Publ. Astron. Soc. Jap.",
    volume = "70",
    number = "SP1",
    pages = "S10",
    year = "2018"
}

@article{Borrow:2022djx,
    author = {Borrow, Josh and Kannan, Rahul and Garaldi, Enrico and Smith, Aaron and Vogelsberger, Mark and Pakmor, R{\"u}diger and Springel, Volker and Hernquist, Lars},
    title = "{thesan-hr: how does reionization impact early galaxy evolution?}",
    eprint = "2212.03255",
    archivePrefix = "arXiv",
    primaryClass = "astro-ph.GA",
    doi = "10.1093/mnras/stad2523",
    journal = "Mon. Not. Roy. Astron. Soc.",
    volume = "525",
    number = "4",
    pages = "5932--5950",
    year = "2023"
}

@ARTICLE{Adamo25_reviewJWST3yrs,
       author = {{Adamo}, Angela and {Atek}, Hakim and {Bagley}, Micaela B. and {Ba{\~n}ados}, Eduardo and {Barrow}, Kirk S.~S. and {Berg}, Danielle A. and {Bezanson}, Rachel and {Brada{\v{c}}}, Maru{\v{s}}a and {Brammer}, Gabriel and {Carnall}, Adam C. and {Chisholm}, John and {Coe}, Dan and {Dayal}, Pratika and {Eisenstein}, Daniel J. and {Eldridge}, Jan J. and {Ferrara}, Andrea and {Fujimoto}, Seiji and {Graaff}, Anna de and {Habouzit}, Melanie and {Hutchison}, Taylor A. and {Kartaltepe}, Jeyhan S. and {Kassin}, Susan A. and {Kriek}, Mariska and {Labb{\'e}}, Ivo and {Maiolino}, Roberto and {Marques-Chaves}, Rui and {Maseda}, Michael V. and {Mason}, Charlotte and {Matthee}, Jorryt and {McQuinn}, Kristen B.~W. and {Meynet}, Georges and {Naidu}, Rohan P. and {Oesch}, Pascal A. and {Pentericci}, Laura and {P{\'e}rez-Gonz{\'a}lez}, Pablo G. and {Rigby}, Jane R. and {Roberts-Borsani}, Guido and {Schaerer}, Daniel and {Shapley}, Alice E. and {Stark}, Daniel P. and {Stiavelli}, Massimo and {Strom}, Allison L. and {Vanzella}, Eros and {Wang}, Feige and {Wilkins}, Stephen M. and {Williams}, Christina C. and {Willott}, Chris J. and {Wylezalek}, Dominika and {Nota}, Antonella},
        title = "{The first billion years according to JWST}",
      journal = {Nature Astronomy},
         year = 2025,
        month = aug,
       volume = {9},
        pages = {1134-1147},
          doi = {10.1038/s41550-025-02624-5},
archivePrefix = {arXiv},
       eprint = {2405.21054},
 primaryClass = {astro-ph.GA},
       adsurl = {https://ui.adsabs.harvard.edu/abs/2025NatAs...9.1134A}
}

@article{Bouwens:2022ojz,
    author = "Bouwens, R. J. and Illingworth, G. D. and Ellis, R. S. and Oesch, P. A. and Stefanon, M.",
    title = "{z {\ensuremath{\sim}} 2{\textendash}9 Galaxies Magnified by the Hubble Frontier Field Clusters. II. Luminosity Functions and Constraints on a Faint-end Turnover}",
    eprint = "2205.11526",
    archivePrefix = "arXiv",
    primaryClass = "astro-ph.GA",
    doi = "10.3847/1538-4357/ac86d1",
    journal = "Astrophys. J.",
    volume = "940",
    number = "1",
    pages = "55",
    year = "2022"
}

@article{Tinker:2010my,
    author = "Tinker, Jeremy L. and Robertson, Brant E. and Kravtsov, Andrey V. and Klypin, Anatoly and Warren, Michael S. and Yepes, Gustavo and Gottlober, Stefan",
    title = "{The Large Scale Bias of Dark Matter Halos: Numerical Calibration and Model Tests}",
    eprint = "1001.3162",
    archivePrefix = "arXiv",
    primaryClass = "astro-ph.CO",
    doi = "10.1088/0004-637X/724/2/878",
    journal = "Astrophys. J.",
    volume = "724",
    pages = "878--886",
    year = "2010"
}

@article{Wang:2026hwx,
    author = "Wang, Kuan and Mu{\~n}oz, Julian B. and Yung, L. Y. Aaron",
    title = "{An Improved Fit for Linear Halo Bias at High Redshift}",
    eprint = "2604.14312",
    archivePrefix = "arXiv",
    primaryClass = "astro-ph.CO",
    month = "4",
    year = "2026"
}

@article{Leo:2018odn,
    author = "Leo, Matteo and Baugh, Carlton M. and Li, Baojiu and Pascoli, Silvia",
    title = "{A new smooth-$k$ space filter approach to calculate halo abundances}",
    eprint = "1801.02547",
    archivePrefix = "arXiv",
    primaryClass = "astro-ph.CO",
    reportNumber = "IPPP-18-1",
    doi = "10.1088/1475-7516/2018/04/010",
    journal = "JCAP",
    volume = "04",
    pages = "010",
    year = "2018"
}

@article{Bohr:2021bdm,
    author = "Bohr, Sebastian and Zavala, Jes{\'u}s and Cyr-Racine, Francis-Yan and Vogelsberger, Mark",
    title = "{The halo mass function and inner structure of ETHOS haloes at high redshift}",
    eprint = "2101.08790",
    archivePrefix = "arXiv",
    primaryClass = "astro-ph.CO",
    doi = "10.1093/mnras/stab1758",
    journal = "Mon. Not. Roy. Astron. Soc.",
    volume = "506",
    number = "1",
    pages = "128--138",
    year = "2021"
}

@article{Barron:2025dys,
    author = "Barron, Jared and Curtin, David and Liu, Hongwan and Mu{\~n}oz, Julian B. and Roy, Sandip",
    title = "{Constraining dark acoustic oscillations with the high-redshift UV luminosity function}",
    eprint = "2512.01998",
    archivePrefix = "arXiv",
    primaryClass = "astro-ph.CO",
    reportNumber = "CERN-TH-2025-200",
    doi = "10.1088/1475-7516/2026/07/026",
    journal = "JCAP",
    volume = "07",
    pages = "026",
    year = "2026"
}

@article{Yung:2023sib,
    author = "Yung, L. Y. Aaron and Somerville, Rachel S. and Nguyen, Tri and Behroozi, Peter and Modi, Chirag and Gardner, Jonathan P.",
    title = "{Characterizing ultra-high-redshift dark matter halo demographics and assembly histories with the gureft simulations}",
    eprint = "2309.14408",
    archivePrefix = "arXiv",
    primaryClass = "astro-ph.CO",
    doi = "10.1093/mnras/stae1188",
    journal = "Mon. Not. Roy. Astron. Soc.",
    volume = "530",
    number = "4",
    pages = "4868--4886",
    year = "2024"
}

@article{Schneider:2013ria,
    author = "Schneider, Aurel and Smith, Robert E. and Reed, Darren",
    title = "{Halo Mass Function and the Free Streaming Scale}",
    eprint = "1303.0839",
    archivePrefix = "arXiv",
    primaryClass = "astro-ph.CO",
    doi = "10.1093/mnras/stt829",
    journal = "Mon. Not. Roy. Astron. Soc.",
    volume = "433",
    pages = "1573",
    year = "2013"
}

@article{Schive:2015kza,
    author = "Schive, Hsi-Yu and Chiueh, Tzihong and Broadhurst, Tom and Huang, Kuan-Wei",
    title = "{Contrasting Galaxy Formation from Quantum Wave Dark Matter, $\psi$DM, with $\Lambda$CDM, using Planck and Hubble Data}",
    eprint = "1508.04621",
    archivePrefix = "arXiv",
    primaryClass = "astro-ph.GA",
    doi = "10.3847/0004-637X/818/1/89",
    journal = "Astrophys. J.",
    volume = "818",
    number = "1",
    pages = "89",
    year = "2016"
}

@article{Benson:2012su,
    author = "Benson, Andrew J. and Farahi, Arya and Cole, Shaun and Moustakas, Leonidas A. and Jenkins, Adrian and Lovell, Mark and Kennedy, Rachel and Helly, John and Frenk, Carlos",
    title = "{Dark Matter Halo Merger Histories Beyond Cold Dark Matter: I - Methods and Application to Warm Dark Matter}",
    eprint = "1209.3018",
    archivePrefix = "arXiv",
    primaryClass = "astro-ph.CO",
    doi = "10.1093/mnras/sts159",
    journal = "Mon. Not. Roy. Astron. Soc.",
    volume = "428",
    pages = "1774",
    year = "2013"
}

@article{Schneider:2018xba,
    author = "Schneider, Aurel",
    title = "{Constraining noncold dark matter models with the global 21-cm signal}",
    eprint = "1805.00021",
    archivePrefix = "arXiv",
    primaryClass = "astro-ph.CO",
    doi = "10.1103/PhysRevD.98.063021",
    journal = "Phys. Rev. D",
    volume = "98",
    number = "6",
    pages = "063021",
    year = "2018"
}

@article{SDSS:2017bih,
    author = "Abolfathi, Bela and others",
    collaboration = "SDSS",
    title = "{The Fourteenth Data Release of the Sloan Digital Sky Survey: First Spectroscopic Data from the Extended Baryon Oscillation Spectroscopic Survey and from the Second Phase of the Apache Point Observatory Galactic Evolution Experiment}",
    eprint = "1707.09322",
    archivePrefix = "arXiv",
    primaryClass = "astro-ph.GA",
    doi = "10.3847/1538-4365/aa9e8a",
    journal = "Astrophys. J. Suppl.",
    volume = "235",
    number = "2",
    pages = "42",
    year = "2018"
}

@article{DES:2017qwj,
    author = "Troxel, M. A. and others",
    collaboration = "DES",
    title = "{Dark Energy Survey Year 1 results: Cosmological constraints from cosmic shear}",
    eprint = "1708.01538",
    archivePrefix = "arXiv",
    primaryClass = "astro-ph.CO",
    reportNumber = "FERMILAB-PUB-17-279-PPD",
    doi = "10.1103/PhysRevD.98.043528",
    journal = "Phys. Rev. D",
    volume = "98",
    number = "4",
    pages = "043528",
    year = "2018"
}

@article{Reid:2009xm,
    author = "Reid, Beth A. and others",
    title = "{Cosmological Constraints from the Clustering of the Sloan Digital Sky Survey DR7 Luminous Red Galaxies}",
    eprint = "0907.1659",
    archivePrefix = "arXiv",
    primaryClass = "astro-ph.CO",
    reportNumber = "FERMILAB-PUB-09-876-A",
    doi = "10.1111/j.1365-2966.2010.16276.x",
    journal = "Mon. Not. Roy. Astron. Soc.",
    volume = "404",
    pages = "60--85",
    year = "2010"
}

@article{Bullock:2017xww,
    author = "Bullock, James S. and Boylan-Kolchin, Michael",
    title = "{Small-Scale Challenges to the $\Lambda$CDM Paradigm}",
    eprint = "1707.04256",
    archivePrefix = "arXiv",
    primaryClass = "astro-ph.CO",
    doi = "10.1146/annurev-astro-091916-055313",
    journal = "Ann. Rev. Astron. Astrophys.",
    volume = "55",
    pages = "343--387",
    year = "2017"
}

@article{Schultz:2014eia,
    author = "Schultz, Christian and O\~norbe, Jose and Abazajian, Kevork N. and Bullock, James S.",
    title = "{The High-$z$ Universe Confronts Warm Dark Matter: Galaxy Counts, Reionization and the Nature of Dark Matter}",
    eprint = "1401.3769",
    archivePrefix = "arXiv",
    primaryClass = "astro-ph.CO",
    doi = "10.1093/mnras/stu976",
    journal = "Mon. Not. Roy. Astron. Soc.",
    volume = "442",
    number = "2",
    pages = "1597--1609",
    year = "2014"
}

@article{Banik:2019smi,
   title={Novel constraints on the particle nature of dark matter from stellar streams},
   volume={2021},
   ISSN={1475-7516},
   url={http://dx.doi.org/10.1088/1475-7516/2021/10/043},
   DOI={10.1088/1475-7516/2021/10/043},
   number={10},
   journal={Journal of Cosmology and Astroparticle Physics},
   publisher={IOP Publishing},
   author={Banik, Nilanjan and Bovy, Jo and Bertone, Gianfranco and Erkal, Denis and de Boer, T.J.L.},
   year={2021},
   month=Oct, pages={043} }

@article{Planck:2018nkj,
    author = "Aghanim, N. and others",
    collaboration = "Planck",
    title = "{Planck 2018 results. I. Overview and the cosmological legacy of Planck}",
    eprint = "1807.06205",
    archivePrefix = "arXiv",
    primaryClass = "astro-ph.CO",
    doi = "10.1051/0004-6361/201833880",
    journal = "Astron. Astrophys.",
    volume = "641",
    pages = "A1",
    year = "2020"
}

@article{Dekel:2013uaa,
    author = "Dekel, A. and Zolotov, A. and Tweed, D. and Cacciato, M. and Ceverino, D. and Primack, J. R.",
    title = "{Toy Models for Galaxy Formation versus Simulations}",
    eprint = "1303.3009",
    archivePrefix = "arXiv",
    primaryClass = "astro-ph.CO",
    doi = "10.1093/mnras/stt1338",
    journal = "Mon. Not. Roy. Astron. Soc.",
    volume = "435",
    pages = "999",
    year = "2013"
}

@article{Hegde:2025mgj,
    author = "Hegde, Sahil and Furlanetto, Steven R.",
    title = "{Efficient semi-analytic modelling of Pop III star formation from Cosmic Dawn to Reionization}",
    eprint = "2507.19581",
    archivePrefix = "arXiv",
    primaryClass = "astro-ph.GA",
    doi = "10.33232/001c.145070",
    month = "7",
    year = "2025"
}

@article{Silk:1997xw,
    author = "Silk, Joseph and Rees, Martin J.",
    title = "{Quasars and galaxy formation}",
    eprint = "astro-ph/9801013",
    archivePrefix = "arXiv",
    journal = "Astron. Astrophys.",
    volume = "331",
    pages = "L1--14",
    year = "1998"
}
% ~
% \newpage
% ~
% \newpage

%%%%%%%%%%%%%%%%%%%%%%%%%%%%%%%%%%%%%%%%%%%%%%%%%%

%%%%%%%%%%%%%%%%% APPENDICES %%%%%%%%%%%%%%%%%%%%%

\appendix

\section{Parameters chosen}
\label{app:parameters}

Table~\ref{tab:parameters} summarizes our choice of fiducial parameters and the range over which they are varied, including alternate analyses (explored in Appendix~\ref{app:robustness}).

\begin{table}[h]
\begin{tabular}{lcc}
Parameter & Range                           & Prior                    \\ \hline
$\log_{10} \epsilon_\star$                                 & $[-3,0]$                        & Flat                     \\
$\log_{10}M_p/M_\odot$                                 & $[10,15]$                       & Flat                     \\
$\alpha_\star$                                             & $[0,3]$                       & Flat                     \\
$\beta_\star$                                              & $[-3, 0]$                       & Flat                     \\
$d\log_{10} \epsilon_\star/dz$                             & $[-2, 2]$                   & Flat             \\
$d\log_{10} M_p/M_\odot/dz$                            & $[-2, 2]$                   & Fixed/Flat$^1$               \\ \hline
$\sigma_{\mathrm {UV}}$                                & $[0.1, 5.0]$                    & Flat                     \\
$d\sigma_{\mathrm {UV}}/d\log_{10}M_h$                 & $[-2, 2]$                   & Flat                     \\
$d\sigma_{\mathrm {UV}}/dz$                            & {$[-2, 2]$} & Flat \\ \hline
$C_{1,\rm dust}$                                       & $[1.1, 2.1]$                     & Fixed/Flat$^2$               \\
$\beta_{0,\rm dust}$                                       & $[-2.6, -1.6]$                     & Fixed/Flat$^2$              
\end{tabular}
\caption{Astrophysical parameters we vary in our analysis, refer to main text for their definition.
(1): In our default run we only vary $\epsilon_\star$ against $z$, as that is sufficient to explain the UVLFs, but we perform alternate runs varying both. 
(2): We fix them at the \citet{Bouwens16xiion} prescription in our fiducial case, but also show a case where we vary them over the prior range shown.
\label{tab:parameters}
}
\end{table}

\section{Selecting the Power Spectrum bins}
\label{app:krange}

\begin{figure}[btp]
    \centering    \includegraphics[width=0.98\linewidth]{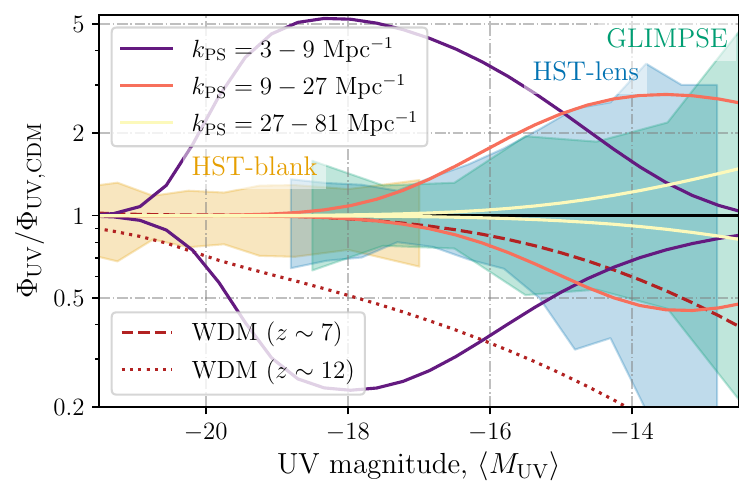}
    \caption{
    Relative change in the UVLF at $z\sim 7$ 
    from multiplying (or dividing) $P(k)$ by 3 within each $k$ bin ($k_{\rm PS}$, each spanning a factor of 3, colored lines), along with the suppression from WDM with $m_{\rm WDM} = 3.2$ keV (dashed; and dotted for $z\sim 12$ where the suppression is stronger). Uncertainties of different UVLF datasets are shown as shaded regions.
    }
    \label{fig:ranges_k_uvlfchange}
\end{figure}

Here we describe our choice of $k$ bins for the matter-power-spectrum analysis. 
We have to determine both reach (i.e., how high on $k$) and resolution (how many bins).
In principle one can follow a principal component analysis to determine how many modes are detectable and at what wavenumbers $k$ they are centered.
Here we follow a simpler approach where we vary the amplitude of the matter power spectrum over different scales to see how it affects the UVLFs.
We choose a factor of 3 ($\approx$ an $e$-fold) of variation on both amplitude and scales.
Fig.~\ref{fig:ranges_k_uvlfchange} shows the result of this test, starting at $k=3-9\,\Mpcinv$ (as $k=1-3\,\Mpcinv$ is 
already well constrained by other probes) where the effect is evident in the abundance of bright galaxies.
For the $k=9-27\,\Mpcinv$ bin we see the effect shift to smaller halos, and therefore fainter galaxies, becoming too small for the HST-blank data (with $\MUV \leq -17$), as found in \citet{Sabti:2021unj}. 
Lensed data (including GLIMPSE) can reach fainter magnitudes, allowing us to measure these scales.
The next bin, with $k = 27-81\, \Mpcinv$ shifts the effect right beyond the reach of current data.
As such, we determine the reach of our data to be $k\approx 27\,\Mpcinv$.

\section{Robustness Checks}
\label{app:robustness}

Here we present our posteriors for the $P(k)$ amplitudes under different assumptions to check their impact, as described in the main text. 
First, we check the effect of the dust prescription, where we vary the two calibration parameters $C_{1,\rm dust}$ and $\beta_{0,\rm dust}$.
Second, we vary $d\log_{10} M_p/dz$ in our MCMC (in the range shown in Tab.~\ref{tab:parameters}).
We also consider only bins brighter than $\MUV = -14$ to test the impact of lensing uncertainties. 
Finally, we do a run where the window function $W(k;M_h)$ is calibrated with $c=2.7$ rather than 2.5. 
In all cases we find results that agree with our fiducial constraint within 1$\sigma$. The farthest deviation is when varying dust, which changes the small-scale amplitude $a_2$ by a half sigma. The uncertainties remain roughly the same, $0.1-0.2$ dex in the first bin and $0.4-0.5$ dex on the second. 

\begin{table}[h]
\centering
\caption{Posteriors on the two $P(k)$ amplitudes under different modeling assumptions, in all cases agreeing within $1\sigma$ with each other.}
\label{tab:amplitudes}
\begin{tabular}{lcc}
\hline
MCMC run & $\log_{10} a_1$ & $\log_{10} a_2$ \\
\hline
Fiducial              & $-0.16^{+0.17}_{-0.16}$ & $-0.19^{+0.45}_{-0.46}$ \\
Varying dust            & $-0.14^{+0.19}_{-0.23}$ & $-0.02^{+0.54}_{-0.56}$ \\
Varying $dM_p/dz$           & $-0.17^{+0.16}_{-0.16}$ & $-0.23^{+0.42}_{-0.47}$ \\
Only $M_{\mathrm{UV}} < -14$ & $-0.16^{+0.17}_{-0.18}$ & $-0.21^{+0.47}_{-0.42}$ \\
With $c=2.7$  &
$-0.17^{+0.16}_{-0.12}$ & $-0.28^{+0.42}_{-0.48}$ \\
\hline
\end{tabular}
\end{table}

\end{document}